\documentclass[twocolumn,secnumarabic,amssymb, nobibnotes, floatfix,aps, prb]{revtex4-1}
\usepackage{graphicx}
\usepackage{epsfig}
\usepackage{amsmath,amssymb,amsfonts}
\usepackage{psfrag}
\usepackage{enumitem}
\usepackage{color}
\usepackage[colorlinks]{hyperref}
\usepackage{epstopdf}
\usepackage{tikz}

\begin{document}

\title{Dynamic pair-breaking current, critical superfluid velocity and nonlinear electromagnetic response of nonequilibrium superconductors}

\author{Ahmad Sheikhzada}
\email{asheikhz@odu.edu}
\author{Alex Gurevich}
\email{gurevich@odu.edu}

\affiliation{Department of Physics and Center 
for Accelerator Science, Old Dominion University, Norfolk, VA 23529, USA}

%\date{\vspace{-5ex}}

\begin{abstract}

We report numerical calculations of a dynamic pairbreaking current density $J_d$ and a critical superfluid velocity $v_d$ in a nonequilibrium superconductor carrying a uniform, large-amplitude ac current density $J(t)=J_a\sin\Omega t$  with $\Omega$ well below the gap frequency $\Omega\ll \Delta_0/\hbar$. The dependencies $J_d(\Omega,T)$ and $v_d(\Omega,T)$ near the critical temperature $T_c$ were calculated from either the full time-dependent nonequilibrium equations for a dirty s-wave superconductor and the time-dependent Ginzburg-Landau (TDGL) equations for a gapped superconductor, taking into account the GL relaxation time of the order parameter $\tau_{GL}$ and the inelastic electron-phonon relaxation time of quasiparticles $\tau_E$. We show that both approaches give similar frequency dependencies of $J_d(\Omega)$ and $v_d(\Omega)$ which gradually increase from their static pairbreaking GL values $J_c$ and $v_c$ at $\Omega\tau_E\ll 1$ to $\sqrt{2}J_c$ and $\sqrt{2}v_c$ at $\Omega\tau_E\gg 1$. Here $J_d$, $v_d$ and a dynamic superheating field at which the Meissner state becomes unstable were calculated in two different regimes of a fixed ac current and a fixed ac superfluid velocity induced by the applied ac magnetic field $H=H_a\sin\Omega t$ in a thin superconducting filament or a type-II  superconductor with a large GL parameter. We also calculated a nonlinear electromagnetic response of a nonequilibrium superconducting state, particularly a dynamic kinetic inductance and a dissipative quasiparticle conductivity, taking into account the oscillatory dynamics of superconducting condensate and the kinetics of quasiparticles driven by a strong ac current. It is shown that an ac current density produces multiple harmonics of the electric field, the amplitudes of the higher-order harmonics diminishing as $\tau_E$ increases.    
\end{abstract}

\maketitle
\section{Introduction}
\label{sec:intro}

Mechanisms of the maximum superfluid velocity $v_c$ and the dc depairing current density $J_c$ which a superconducor can carry in an equilibrium state have been well established ~\cite{tinkh}. The first calculations ~\cite{VL} of $v_c(T)$ and $J_c(T)$ were based on the Ginzburg-Landau (GL) equations near the critical temperature $T\approx T_c$. Furthermore, $v_c(T)$ and $J_c(T)$ have been calculated in the whole temperature range $0<T<T_c$ in the BCS model for clean ~\cite{parment,bardeen,maki1,maki2} and dirty ~\cite{maki1,maki2} superconductors with nonmagnetic and magnetic impurities \cite{kupr} and taking into account strong electron-phonon coupling in the Eliashberg theory ~\cite{nicole}. The dc depairing current densities  have been measured for different superconducting materials ~\cite{jd1,jd2,jd3}. These issues are closely related to a maximum superheating magnetic field $H_s$ which can be sustained by a superconductor in the vortex-free Meissner state. Here $H_s(T)$ near $T_c$ has been calculated from the GL theory ~\cite{matricon,chapman} and for type-II superconductorts with a large GL parameter $\kappa\gg  1$  at $T=0$  ~\cite{galaiko} and in the entire temperature range $0<T<T_c$ both in the clean limit ~\cite{catelani} and for arbitrary concentrations of nonmagnetic and magnetic impurities ~\cite{lin}.  Nonlinear screening and breakdown of superconductivity in proximity-coupled bilayers under a strong dc magnetic field have been calculated in Refs. \onlinecite{ns1,ns2,ns3,ns4} . 

Unlike the static $v_c$ and $J_c$ in equilibrium, the physics of the dynamic critical superfluid velocity $v_d$ and the depairing current density $J_d$ at which superconductivity is destroyed in a {\it nonequilibrium} state is not well understood. The dynamic $v_d$ and $J_d$ are controlled by both the nonlinear current pairbreaking effects and a complex kinetics of quasiparticles driven out of equilibrium by a time-dependent electromagnetic field ~\cite{kopnin}. For an oscillating superflow  $J(t)=J_a\sin\Omega t$, the dynamic $v_d$ and $J_d$ depend on the frequency $\Omega$ and the relaxation time constants for the superfluid density $\tau_{GL}(T)$ and quasiparticles $\tau_E(T)$. At $\Omega\ll\Delta/\hbar$ the ac field does not generate new quasiparticles which transfer the absorbed power to phonons.  At $k_BT\ll\Delta$ this power transfer is mostly limited by an inelastic scattering time of quasiparticles $\tau_s(T)$ and a  recombination time of Cooper pairs $\tau_r(T)$ due to electron-phonon collisions  ~\cite{kaplan}:
\begin{eqnarray}
\tau_r \simeq \tau_1 \biggl( \frac{T_c}{T} \biggr)^{1/2} e^{\Delta/T}, 
\hspace{1cm}
\tau_s \simeq \tau_2 \biggl( \frac{T_c}{T} \biggr)^{7/2}, 
\label{tauu}
\end{eqnarray}
where $\tau_1$ and $\tau_2$ are materials constants.  Depending on the amplitude $J_a$, the distribution function of quasiparticles $f(E,t)$ can either deviate strongly from the Fermi-Dirac distribution $f_0(E)$ at $(\tau_r,\tau_s)\Omega \gg 1$  or relax to $f_0(E)$ at $ (\tau_r,\tau_s)\Omega\ll 1$.  Since both $\tau_r(T)$ and $\tau_s(T)$ increase as $T$ decreases,  nonequilibrium effects become more pronounced at $T\ll T_c$. By contrast, $\tau_{GL}(T)$ increases as $T$ increases and diverges at $T=T_c$ ~\cite{kopnin}
\begin{equation}
\tau_{GL}(T)\simeq \frac{\pi\hbar}{8k_B(T_c-T)},\qquad T\approx T_c.
\end{equation}
At $T\ll T_c$ the condition $\Omega\tau_{GL}\lesssim 1$ is satisfied up to $0.1-1$ THz  for most superconductors but breaks down at temperatures very close to $T_c$. For instance, at 1 GHz, we have $\Omega\tau_{GL}(T)\simeq 1$ at $T_c-T\simeq \pi\hbar\Omega/8k_B\sim 10^{-2}$K.  

The dynamics of the condensate at $\Omega\tau_{GL}\ll 1$ remains nearly quasistatic if the effect of quasiparticles is weak. At $T\ll T_c$, the relaxation times $\tau_s(T)$ and $\tau_r(T)$ increase strongly as the temperature decreases so that $(\tau_r,\tau_s)\Omega\gtrsim 1$  while $\Omega\tau_{GL} \ll 1$, and the ac field can produce highly nonequilibrium quasiparticles.  Yet the density of quasiparticles in s-wave superconductors  at $T\ll T_c$ and $\Omega\ll \Delta/\hbar$ is exponentially small as compared to the superfluid density, so the nonequilibrium quasiparticles have only a weak effect on the dynamics of the condensate which reacts almost instantaneously to $J(t)$. In this case, the dynamic $v_d$ and $J_d$ at $\Omega\ll \Delta/\hbar$ and $T\ll T_c$ would be close to the static $v_c$ and $J_c$ in thermodynamic equilibrium.

The situation changes at $T\approx T_c$ where the superfluid density becomes smaller than the density of nonequilibrium quasiparticles which significantly affect the dynamic $v_d$ and $J_d$ at which superconductivity breaks down.  In this work we used both the time-dependent Ginzburg-Landau (TDGL) equations and a full set of nonequilibrium  equations for dirty superconductors in a low-frequency $(\Omega\ll \Delta/\hbar)$ field ~\cite{kopnin,ss,LO,Kr1,Kr2} to calculate the dynamic $v_d(T,\Omega)$ and $J_d(T,\Omega)$  at $T\simeq T_c$, where nonequilibrium effects are most pronounced. We consider the case of $\hbar\Omega\ll k_BT$ in which the microwave stimulation of superconductivity ~\cite{eliashberg} does not happen, but the ac currents strongly affect the density of states of quasiparticles ~\cite{maki2,fulde,denscur} and drive them out of equilibrium.   

The physics of the dynamic critical velocity is relevant to many applications, for instance, microwave thin film superconducting resonators used in kinetic inductance photon detectors and astrophysical spectroscopy\cite{kid,caltech}. It is also essential for superconducting resonant cavities for particle accelerators, where the breakdown fields close to the thermodynamic superheating field $H_s$ have been achieved at very high quality factors $\sim 10^{10}$ at 2K in the Meissner state ~\cite{Padamsee,ag_srf}.  These cavities operate at $0.1-3$ GHz much lower than the gap frequency $\Delta/h\simeq 0.8$ THz for Nb, and the dynamic superheating field $H_d$ sets a theoretical limit of the rf breakdown. The dynamic superheating field was measured by Yogi et al. \cite{q1} who showed that for Sn, Pb, In at 90-300 MHz, the breakdown field near $T_c$ is close to $H_s(T)$. Pulse measurements ~\cite{q2} on Nb and Nb$_3$Sn at GHz frequencies at $2$K$ <T<T_c$ have shown that the field onset of magnetic flux penetration is close to $H_s(T)$ for Nb near $T_c$ but is smaller than $H_s(T)$ for Nb$_3$Sn at lower $T$. 

In this work we calculate the dynamic $J_d(\Omega,T)$ and a critical phase gradient $Q_d(\Omega,T)$ of the order parameter related to $v_d$ by $Q_d=mv_d/\hbar$, where $m$ is the electron mass \cite{tinkh} for a uniform ac superflow at $T\simeq  T_c$.   We focus here on the maximum amplitude of the ac current density $J(t)=J_a\sin\Omega t$ which can be sustained in a nonequilibrium Meissner states and do not consider nonuniform dissipative states at $J_a>J_d$ due to proliferation of phase slip centers in narrow filaments~\cite{ps1,ps2,ps3} or penetration of vortices in bulk superconductors above the dynamic superheating field.    
TDGL simulations of thin filaments have shown that $J_d$ can approach $\sqrt{2}J_c$ at $\Omega\tau_E\gg 1$ ~\cite{ps3}, while numerical simulations of kinetic equations ~\cite{Kr1,Kr2} have shown \cite{VP1} that superconductivity can persist during short current pulses with amplitudes above the static $J_c$. Yet the calculations of $J_d$ and $Q_d$ taking into account both the nonlinear current pairbreaking and nonequilibrium kinetics of quasiparticles, have not yet been done.   We also calculate a nonlinear electromagnetic response in a nonequilibrium state at $J<J_d$ and its manifestations in the nonlinear Meissner effect, kinetic inductance and intermodulation which have been so far investigated in equilibrium superconductors ~\cite{Yip,Dahm,Anlage,Hirsch,Oates,Groll,kind1,kind2,kind3,kind4,kind5}.    

The paper is organized as follows. In Sec. \ref{sec:eq} we specify the main equations and discuss the theoretical assumptions under which the equations have been derived. These equations were solved for a uniform ac superflow in Sec. \ref{sec:qc}, where the dynamic $Q_d(T,\Omega)$ and $J_d(T,\Omega)$ were calculated. In Sec. \ref{sec:man} we address a nonlinear response and calculate the current-dependent kinetic inductance both in  equilibrium and nonequilibrium states. The conclusions and broader implications of our results are presented in Sec. \ref{sec:disc}.

\section{Main Equations}
\label{sec:eq}

\begin{figure}
\vspace{-30mm}
\includegraphics[width=\columnwidth]{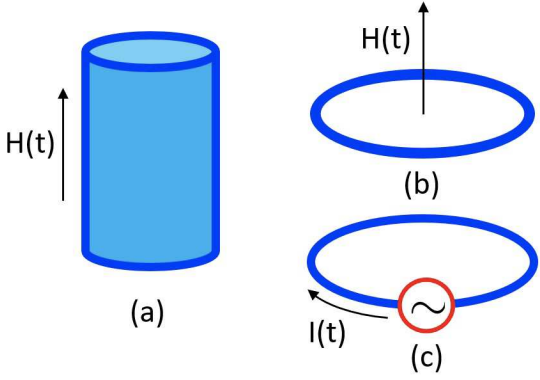}
\vspace{-30mm}
\caption{Geometries for which $Q_d(\Omega,T)$ and $J_d(\Omega,T)$ are calculated:  (a) a thin film cylinder in a parallel ac magnetic field, (b) a thin filament ring in a perpendicular magnetic field, (c) a thin wire connected to an ac power supply. }
\label{fig1}
\end{figure}

We consider a dirty s-wave superconductor exposed to time-dependent electromagnetic potentials $\mathbf{A}(\mathbf{r},t)$ and $\varphi(\mathbf{r},t)$. The dynamic $Q_d(\Omega,T)$ and $J_d(\Omega,T)$ at $T\simeq T_c$ are calculated using the equations for the order parameter $\Psi(\mathbf{r},t)=\Delta\exp(-i\theta)$ and the current density $\mathbf{J}(\mathbf{r},t)$ along with a kinetic equation for the distribution function of quasiparticles ~\cite{ss,LO,Kr1,Kr2}. The cases of a fixed ac  superfluid velocity $v(t)$ and a fixed ac current density $J(t)$ are investigated. These cases can be realized in the geometries shown in Fig. \ref{fig1}, where a thin film cylinder (a) and a ring filament (b) exposed to the ac magnetic field $H(t)$ correspond to the regime of fixed $v(t)$, whereas a thin wire connected to an ac power supply shown in Fig. \ref{fig1} (c) or a semi-infinite superconductor with $\kappa\gg 1$ corresponds to the regime of fixed $J(t)$. It is assumed that the thickness $d$ of films and filaments is much smaller than the magnetic penetration depth $\lambda_L$, so that the induced current density is uniform over the cross-section.  We focus here on the stability of a uniform Meissner state and do not consider thermally-activated or quantum proliferation of vortices or phase-slip centers ~ \cite{aps1,aps2,qps1,qps2} and the influence of ac current  ~\cite{psa1,psa2} on their dynamics at $J<J_d(\Omega,T)$, or the effects of inhomogeneities \cite{psinh} and current leads on the nucleation of vortices or phase slips. The condition that vortices do not nucleate at $J\sim J_d$ requires $d\lesssim \xi(T)$, where $\xi$ is the coherence length. It is also assumed that the magnetic flux threading the samples shown in Fig. \ref{fig1}a is much greater than the flux quantum $\phi_0$ and the Little-Parks oscillations ~\cite{tinkh} are washed out. Here the self field is smaller than the applied field by the factor $d/\lambda_L\ll 1$.

The dynamic $Q_d(\Omega,T)$ and $J_d(\Omega,T)$ for both fixed electric field and fixed current are calculated by first solving the TDGL equations. The TDGL approach is useful to address qualitative mechanisms of destruction of superconductivity by an ac current, even though the TDGL theory, strictly speaking, is not  applicable for the calculations of $J_d(\Omega,T)$.  We then calculate $Q_d(\Omega,T)$ and $J_d(\Omega,T)$ by solving the full set of dynamic equations of Ref. \onlinecite{LO}. Comparing the TDGL results with a more adequate theory of Refs. \onlinecite{ss,LO,Kr1,Kr2} shows the effects of nonequilibrium kinetics of quasiparticles and the extent to which the TDGL approach is applicable. We then proceed with the calculations of the kinetic inductance and the nonlinear electromagnetic response in nonequilibrium states. 

\subsection{TDGL equations}
\label{sec:gtdgl}
Slow temporal and spatial variations of $\Psi(\mathbf{r},t)$ and $\mathbf{J}(\mathbf{r},t)$ in a dirty s-wave superconductor at $T\approx T_c$ can be described by the TDGL equations ~\cite{Kr1,Kr2}:
\begin{gather}
\frac{\pi}{8T_c\epsilon}(1+4\tau_E^2\Delta^2)^{-1/2}\left(\frac{\partial}{\partial t}+2ie\varphi+2\tau_E^2\frac{\partial \Delta^2}{\partial t}\right)\Psi \nonumber\\ =\left(1-\frac{\Delta^2}{\Delta_0^2}\right)\Psi+\xi^2\left(\mathbf{\nabla}-2ie\mathbf{A}\right)^2\Psi,
\label{gtdgl}\\
\mathbf{J}=\frac{\pi\sigma_0}{4eT_c}\Delta^2\mathbf{Q}-\sigma_0\left(\mathbf{\nabla}\varphi+\frac{\partial \mathbf{A}}{\partial t}\right).
\label{j}
\end{gather}
Here $\xi=(\pi \hbar D/8k_BT_c\epsilon)^{1/2}$ is the coherence length, $D=v_F l/3$ is diffusion constant, $v_F$ is the Fermi velocity, $l$ is the mean free path, $\epsilon=1-T/T_c$, $\tau_E$ is an energy relaxation time due to inelastic scattering of quasiparticles on phonons \cite{kopnin}, 
$\Delta_0^2=8\pi^2k_B^2T_c^2\epsilon/7\zeta(3)$, $\sigma_0=2e^2DN(0)$ is the normal state conductivity, $N(0)$ is the density of states on the Fermi surface, $-e$ is the electron charge, and $\mathbf{Q}=-(\nabla\theta+2\pi\mathbf{A}/\phi_0)$ is a gauge-invariant phase gradient. Equations (\ref{gtdgl}) and (\ref{j}) (in which the units with $\hbar=k_B=1$ are used) were derived from the kinetic BCS  theory under the condition of local equilibrium, assuming that $\mathbf{Q}(\mathbf{r},t)$ and $\Delta(\mathbf{r},t)$ vary slowly over $\xi_0$, the diffusion length $L_E=(D\tau_E)^{1/2}$ and $\tau_E$ ~\cite{Kr1,Kr2,kopnin}, where   
\begin{equation}
\tau_E=\frac{8\hbar}{7\pi\zeta(3)\lambda k_B T}\left(\frac{c_s}{v_F}\right)^2\left(\frac{T_F}{T}\right)^2.
\label{tau}
\end{equation}
Here $c_s$ is the speed of longitudinal sound, $\lambda$ is a dimensionless electron-phonon coupling constant, and $T_F=\epsilon_F/k_B$ is the Fermi temperature. For Pb, we have ~\cite{carbotte,ashkroft} $c_s\simeq 1.32$ km/s, $v_F\simeq 1830$ km/s, $T_F=1.1\cdot 10^5$ K, $T_c=7.3$ K and $\lambda=1.55$, which yields $\tau_E^{Pb}(T_c)\simeq 2.52\cdot 10^{-11}$s. For Al with $c_s\simeq 5.1$ km/s, $v_F\simeq 2030$ km/s, $T_F=1.36\cdot 10^5$ K, $T_c=1.2$ K and $\lambda=0.43$, Eq. (\ref{tau})  gives $\tau_E^{Al}(T_c)\simeq 3.64\cdot 10^{-7}$ s. 
 
For a uniform superflow, Eqs. (\ref{gtdgl}) and (\ref{j}) in the gauge $\varphi=0$ can be written in the following dimensionless form: 
\begin{gather}
(1+4\tau^2\psi^2)^{1/2}\frac{\partial \psi}{\partial t}=(1-q^2)\psi-\psi^3,
\label{glq}\\
j = u\psi^2q+\frac{\partial q}{\partial t},
\label{glj}
\end{gather} 
where $\psi=\Delta/\Delta_0$, $q=Q\xi$, $\tau=\Delta_0\tau_E/\hbar$, $j=J/J_0$, $t$ is in units of $\tau_{GL}$, $J_0=\sigma_0/2e\xi\tau_{GL}$,  and $u=\pi^4/14\zeta(3)\approx 5.79$.

\subsection{Nonequilibrium kinetic equations}
\label{sec:noneq}

For a uniform current flow, the full set of nonequilibrium kinetic equations ~\cite{LO,Kr1,Kr2} given in Appendix \ref{Ap1} can be reduced to a single kinetic equation for the odd in energy $E$ part of the quasiparticle distribution function $f(E,t)$, and dynamic equations for $\psi(t)$ and $j(t)$:
\begin{gather}
R_{2}\frac{\partial f}{\partial E}\frac{\partial\psi}{\partial t}+N_{1}\left(\frac{\partial}{\partial t}+\frac{s}{2\tau}\right)\delta f=
\frac{N_{2}R_{2}}{s}\frac{\partial f}{\partial E}\frac{\partial q^2}{\partial t},
\label{k1}
\\
\frac{\partial\psi}{\partial t}-\frac{1}{\epsilon}\int_{0}^{\infty}R_2\delta f dE=\left(1-q^{2}\right)\psi-\psi^3,
\label{gap}
\\
j=u\psi^2q+\frac{\partial q}{\partial t}\int_0^\infty (N_1^2+N_2^2)\frac{\partial f}{\partial E}dE+\nonumber \\
2qs\int_0^\infty N_2R_2\delta f dE,\qquad s=(u/\epsilon)^{1/2}.
\label{jeq}
\end{gather}
Here $\delta f(E,t)=f(E,t)-f_0(E)$, $f_0=\tanh(E/2T)$, the quasiparticle energy $E$ and temperature $T$ are in units of $\Delta_0$, and the scaling factor $(u/\epsilon)^{1/2}=2\tau_{GL}\Delta_0/\hbar$ results from the same normalization of the parameters as in Eqs. (\ref{glq}) and (\ref{glj}). If $\Omega\tau_{GL} \ll 1$ the spectral functions $N_1,N_2,R_1$ and $R_2$ are defined by the normal $\alpha(E)=N_1(E)+iR_1(E)$ and anomalous $\beta(E)=N_2(E)+iR_2(E)$ Green's functions which satisfy the quasi-static Usadel equation for 1D current flow ~\cite{Kr1,Kr2}:
\begin{equation}
\left(\frac{1}{2\tau}-iE\right)\beta+\frac{q^2}{2}\alpha\beta=\psi\alpha,
\label{usad}
\end{equation}
where $\alpha^2+\beta^2=1$. Eq. (\ref{usad}) reduces to a quatric equation for $\alpha$, the solutions of which are given in Appendix \ref{Ap1}. The term $1/2\tau$ in Eq. (\ref{usad}) defines a finite quasiparticle lifetime due to scattering on phonons, resulting in subgap states at $|E|<\psi$. We do not consider here other contributions to the subgap states  ~\cite{dynes,JohnZ,kg}.

We solved the integro-differential Eqs. (\ref{k1})-(\ref{jeq}) numerically using the method of lines ~\cite{mdln}. By discretizing the energy, Eqs. (\ref{k1})-(\ref{jeq}) were reduced to coupled ordinary differential equations in time which were solved by the Adams-Bashforth-Moulton method ~\cite{mdabm} with the error tolerances below $10^{-6}$.  Results of the calculations of the dimensionless $j_d=J_d/J_0$ and $q_d=Q_d\xi$ as functions of the dimensionless frequency $\omega=\Omega\tau_{GL}$ and the quasiparticle relaxation time $\tau=\tau_E\Delta_0/\hbar$ are given below.
 
\section{Dynamic pairbreaking current }
\label{sec:qc} 
\subsection{TDGL results}
\label{sec:Glqc}

The stationary Eqs. (\ref{glq})-(\ref{glj}) have the solution $\psi=0$ at $q>1$ and $\psi=\sqrt{1-q^2}$ at $q<1$. Stability of this solution with respect to small perturbations $\delta\psi(t)$ and $\delta q(t)$ depends on the way by which the superflow is generated.  In the regime of fixed $q$ the stationary solution  $\psi(q)$ is stable in the whole region of $q<q_c=1$, but in the regime of fixed $j$ the solution $\psi(q)$ is stable if $q$ is smaller than $q_c=1/\sqrt{3}$ at which $j=uq(1-q^2)$ reaches maximum \cite{tinkh,VL}. This gives the GL depairing current density $j_c=2u/3\sqrt{3}$ above which $\psi(j)$ drops from $\psi(j_c)=\sqrt{2/3}$ to zero. 

\subsubsection{Fixed Q(t).}
\label{fq}

\begin{figure}
\includegraphics[width=\columnwidth ]{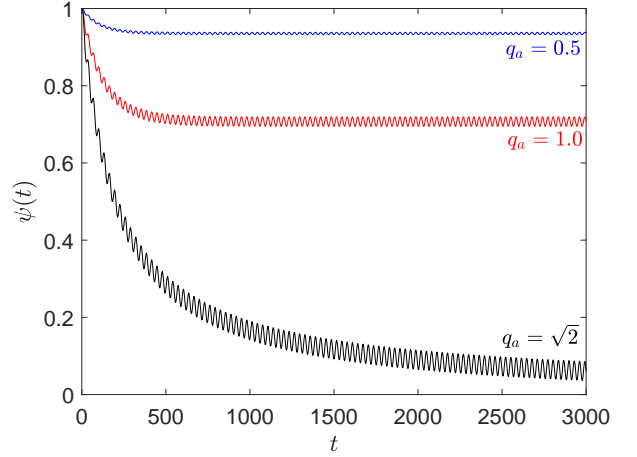}
\caption{Dynamics of $\psi(t)$ calculated at $q=q_a\sin \omega t$, $\tau=100$, and $\omega = 0.1$. Here $\psi(t)$ eventually vanishes at $q_a=\sqrt{2}$.}
\label{fig2}
\end{figure}

Figure \ref{fig2} shows  $\psi(t)$ calculated from Eq. (\ref{glq}) with $q(t)=q_a\sin\omega t$ at $\omega=\Omega\tau_{GL}=0.1$, $\tau=100$ and the initial condition $\psi(0)=1$.  Here $\psi(t)$ relaxes after a transient period $t\gtrsim \sqrt{1+4\tau^2}$ to an oscillating steady-state with a nonzero mean $\langle\psi\rangle$ if $q_a<q_d(\omega,T)$ or to the normal state with $\psi(t)=0$ at $t\gg 1$ if $q_a>q_d(\omega,T)$. The mean $\langle\psi(q_a)\rangle$ decreases with $q_a$ and vanishes at $q_a=q_d$.

\begin{figure}
\includegraphics[width=\columnwidth ]{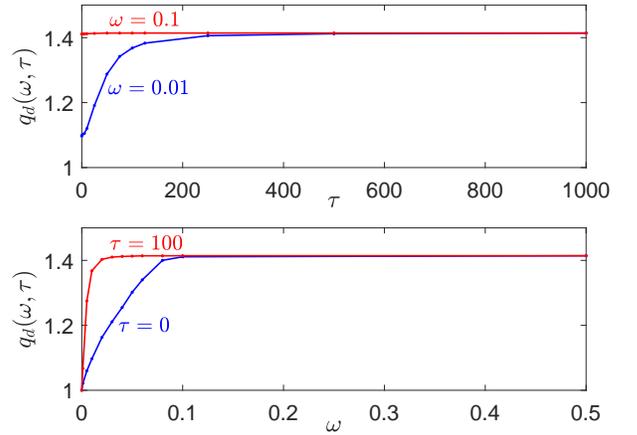}
\caption{The calculated dependencies of $q_d$ on $\tau$ (top) and $\omega$ (bottom). Here $q_d\to\sqrt{2}$ at $\omega\tau\gtrsim 1$. }
\label{fig3}
\end{figure}

\begin{figure}
\includegraphics[width=\columnwidth ]{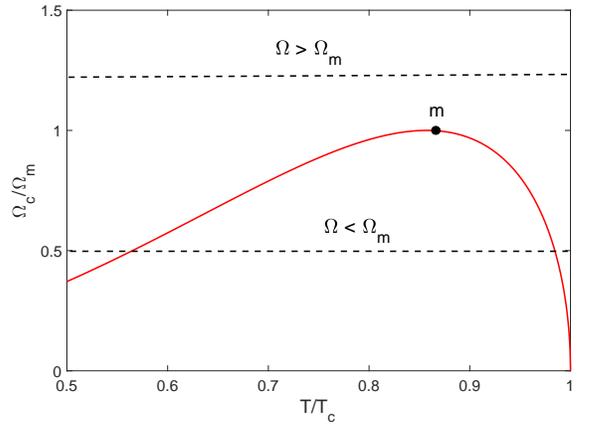}
\caption{Temperature dependence of $\Omega_c(T)$. The dashed lines show the levels of fixed $\Omega$ at $\Omega>\Omega_m$ and $\Omega<\Omega_m$, where 
$\Omega_m$ is the maximum value of $\Omega_c(T)$ corresponding to the point $m$.}
\label{fig4}
\end{figure}

The calculated dependencies of $q_d$ on $\omega$ and $\tau$ are shown in Fig. \ref{fig3}. Here $q_d(\tau)$ at $\omega=0.01$ increases from $q_d(0)\approx 1.097$ at $\tau=0$ to $q_d(\tau)\to \sqrt{2}$ at $\tau\gg 1$. At higher frequency $\omega=0.1$, the dynamic $q_d(\tau)$ is nearly equal to $\sqrt{2}$ at all $\tau$.  However, if $\tau$ is fixed but the frequency changes, $q_d(\omega)$ varies from $q_c=1$ at $\omega=0$ to $q_d(\omega)\to \sqrt{2}$ at $\omega\sqrt{1+4\tau^2}\gg 1$ . The universal value of $q_d=\sqrt{2}$ is achieved at $\omega\tau\gtrsim 1$, that is, for $\Omega$ exceeding a crossover frequency $\Omega_c \simeq \hbar/\tau_{GL}\Delta_0\tau_E$ given by:
\begin{equation}
\Omega_c \simeq \frac{ k_B}{\Delta_0\tau_E}(T_c-T)\sim\frac{k_BT^3}{\hbar T_D^2}\sqrt{1-\frac{T}{T_c}},
\label{omc}
\end{equation}
where $T_D$ is the Debye temperature. Here $\Omega_c(T)$ vanishes at $T_c$, reaches maximum $\Omega_m=\Omega_c(6T_c/7)$ at $T/T_c\approx 0.86$ and decreases with $T$ at $T < 0.8 T_c$, as shown in Fig. \ref{fig4}.

The increase of $Q_d(\Omega,T)$ at $\Omega\gtrsim \Omega_c(T)$ by the factor $\sqrt{2}$ can be understood as follows. As follows from Fig. \ref{fig2}, $\psi(t)$ oscillates rapidly around a mean $\langle\psi\rangle$. Here $\langle\psi\rangle\simeq \sqrt{1-\langle q^2\rangle}$ is determined by Eq. (\ref{glq}) with the time-averaged $\langle q^2(t)\rangle=q_a^2/2$ so $\langle\psi\rangle$ vanishes at $q_a=\sqrt{2}$. A small-amplitude ac correction $\delta\psi(t)$ was calculated in Appendix \ref{Ap2}.  The superconducting state remains stable in the whole region $0<q_a<q_d$.

\begin{figure}
\includegraphics[width=\columnwidth ]{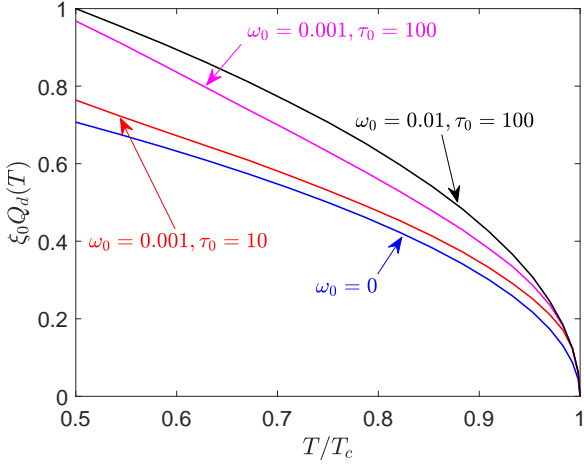}
\caption{$Q_d(T)$ calculated for different values of $\omega_0=\pi\hbar\Omega/8k_BT_c$ and $\tau_0=\tau_E(T_c)\Delta_0(0)$, where $\Delta_0^2(0)=8\pi^2T_c^2/7\zeta(3)$. Here the dynamic $Q_d=\sqrt{2(1-T/T_c)}/\xi_0$ at $\Omega\gg \Omega_c(T)$ has the same temperature dependence as the static $Q_c=\sqrt{1-T/T_c}/\xi_0$. If $\Omega\sim\Omega_c(T)$ the behavior of $Q_d(T)$ is affected by the temperature dependence of $\tau_E(T)$, as shown for the case of $\omega_0=0.001$ and $\tau_0=100$. }
\label{fig5}
\end{figure}

The temperature dependence of $Q_d(\Omega,T)$ shown in Fig. \ref{fig5} is affected by the ratio $\Omega/\Omega_c(T)$. If $\Omega>\Omega_m=\Omega_c(6T_c/7)$ (see Fig. \ref{fig4}), the dynamic $Q_d(T)\to\sqrt{2}/\xi(T)$ has the same temperature dependence as the static $Q_c=1/\xi(T)$. However, if $\Omega \ll \Omega_m$, we obtain that $Q_d(T)\to \xi_0^{-1}\sqrt{2(1-T/T_c)}$ at $T$  close to $T_c$ and crosses over to the static $Q_c(T)$ at lower $T$. There is also a range of frequencies $\Omega < \Omega_m$ but $\Omega\gtrsim \Omega_c(T_c/2)$ (see Fig. \ref{fig4}) in which $Q_d(T)$ evolves from $\sqrt{2}Q_c(T)$ at $T\to T_c$ to $Q_d\simeq Q_c(T)$ at $T\lesssim 0.8T_c$ and back to $\simeq \sqrt{2}Q_c(T)$. 

\subsubsection{Fixed J(t).}

\begin{figure}
\includegraphics[width=\columnwidth ]{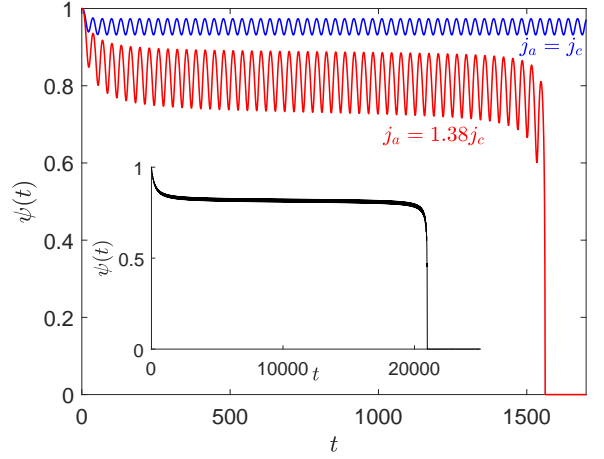}
\caption{Dynamics of $\psi(t)$ calculated at $j=j_a\sin \omega t$, $\omega=0.1$, $\tau=10$, and different amplitudes $j_a$. At $j_a=j_c$, the superconducting state still exists, but once $j_a$ reaches the dynamic pair breaking current $j_d=1.38j_c$,  $\psi(t)$ vanishes. The inset shows $\psi(t)$ calculated at $\tau=100$ at $j_a=\sqrt{2}j_c$ and $\omega=0.1$. }
\label{fig6}
\end{figure}

\begin{figure}
\includegraphics[width=\columnwidth ]{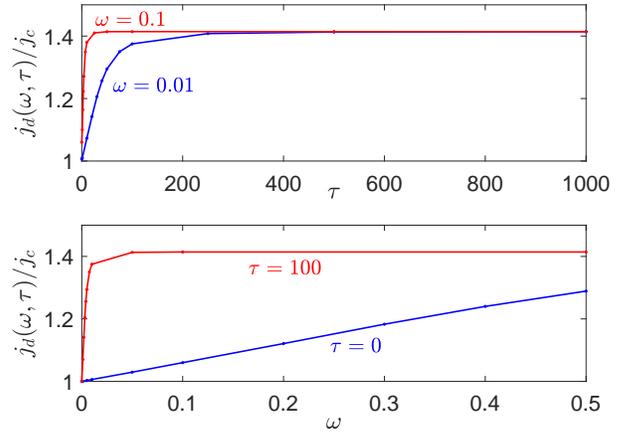}
\caption{Dynamic pair breaking current $j_d$ as a function of $\tau$ (top) and $\omega$ (bottom). Here $j_d(\omega,\tau)\to \sqrt{2}j_c$ at $\omega\tau\gg 1$. }
\label{fig7}
\end{figure}

We calculated $\psi(t)$ at a fixed $j(t)=j_a\sin\omega t$ by solving the coupled Eqs. (\ref{glq})-(\ref{glj}). The GL dc depairing current density $j_c=2u/3\sqrt{3}\approx 2.228$ is reached at $q=1/\sqrt{3}$ and $\psi^2=2/3$, while at $q>1/\sqrt{3}$  the superconducting state becomes unstable and $\psi(q)$ vanishes abruptly \cite{tinkh}. This feature is characteristic of the ac current as well, which makes it different from the regime of fixed $q(t)$. For instance, Fig. \ref{fig6} shows $\psi(t)$ calculated at $\tau=10$ and $\omega=0.1$. At $j_a=1.38j_c$ the order parameter abruptly vanishes after a transient period. For large $\tau$, this transition to the normal state occurs at $j_a=\sqrt{2}j_c$, as shown in the inset for $\tau=100$ and $\omega=0.1$. Here the dynamic pair breaking current $j_d(\omega,\tau)$ shown in Fig. \ref{fig7} exhibits similar dependencies on $\omega$ and $\tau$ as $q_d(\omega,\tau)$ at a fixed $q(t)$. If $\omega\tau\gtrsim 1$ both the dynamic $j_d(\omega,\tau)$ and $q_d(\omega,\tau)$ are larger by the factor $\sqrt{2}$ than their respective GL values. 

\begin{figure}
\includegraphics[width=\columnwidth ]{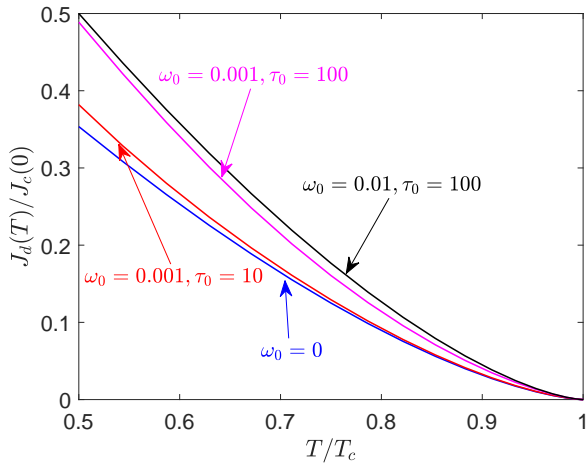}
\caption{$J_d(T)$ calculated for different values of $\omega_0=\pi\hbar\Omega/8k_BT_c$ and $\tau_0=\tau_E(T_c)\Delta_0(0)$, where $\Delta_0^2(0)=8\pi^2T_c^2/7\zeta(3)$. Here the dynamic $J_d=\sqrt{2}J_c(0)(1-T/T_c)^{3/2}$ at $\Omega\gg \Omega_c(T)$ has the same temperature dependence as the static $J_c=J_c(0)(1-T/T_c)^{3/2}$. At $\Omega\sim\Omega_c(T)$ the behavior of $J_d(T)$ is affected by the temperature dependence of $\tau_E(T)$, as shown for the case of $\omega_0=0.001$ and $\tau_0=100$.}
\label{fig8}
\end{figure}

The temperature dependence of $J_d(\Omega,T)$ is affected by the temperature dependencies of $\tau(T)$ and $\Omega_c(T)$. At $T \to T_c$ and 
$\Omega\gtrsim\Omega_c(T)$ the dynamic pair breaking current $J_d$ is $\sqrt{2}$ times larger than the static $J_c(T)$ and is independent of $\tau$. As $T$ decreases $J_d(\Omega,T)$ can evolve to $J_c(T)$ at temperatures for which $\Omega\lesssim \Omega_c(T)$. This behavior of $J_d(\Omega,T)$ is illustrated in Fig. \ref{fig8}. 

\subsection{$Q_d(T,\Omega)$ and $J_d(T,\Omega)$ calculated from the full set of nonequilibrium equations}
\label{sec:nonqc}

The TDGL calculations of $q_d(T,\omega)$ and $j_d(T,\omega)$ give a qualitative picture of dynamic pairbreaking, although Eqs. (\ref{glq})-(\ref{glj}) are not really applicable at $J\simeq J_d$. Indeed, the dynamic terms in Eqs. (\ref{glq})-(\ref{glj}) were derived from the BCS kinetic theory, assuming weak pairbreaking and local equilibrium in which  $Q\xi\ll 1$ and $\Delta(\mathbf{r},t)$ varies slowly over the diffusion length $L_E=(D\tau_E)^{1/2}$ and the energy relaxation time $\tau_E$~ \cite{Kr1,Kr2}.  Those conditions break down at $Q\simeq Q_c\sim \xi^{-1}$ and $\Omega\gtrsim \tau_{GL}^{-1}$, so in this section we calculate  $\psi(t)$, $q_d(T,\omega)$ and $j_d(T,\omega)$ from Eqs. (\ref{k1})-(\ref{jeq}) which take into account both the dynamic current pairbreaking and nonequilibrium kinetics of quasiparticles. 

\begin{figure}
\includegraphics[width=\columnwidth]{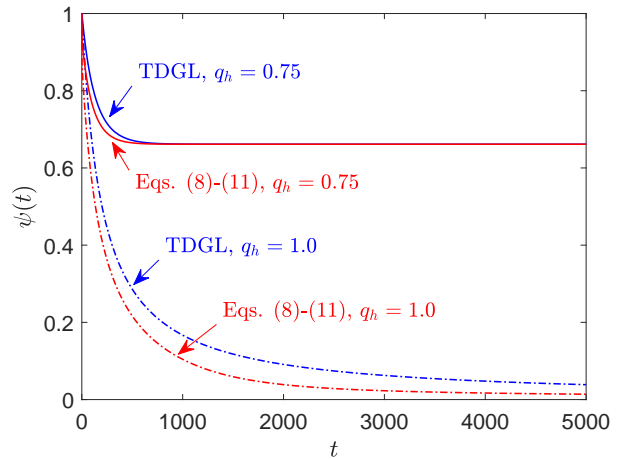}
\caption{ Comparison of $\psi(t)$ calculated from the TDGL equation (\ref{glq}) and the full nonequilibrium Eqs. (\ref{k1})-(\ref{usad}) for $q(t)=q_h\tanh t$ at $q_h=0.75$ and $q_h=1$. Here we took $\tau(T)=100$ and $T=0.9T_c$.  }
\label{fig9}
\end{figure}

Consider first solutions of Eqs. (\ref{k1})-(\ref{usad}) at $\tau(T) = 100$ and $T=0.9T_c$ for a superflow $q(t)=q_h\tanh t$ which was gradually turned on at $t=0$. As shown in Fig. \ref{fig9}, the qualitative behavior of $\psi(t)$ calculated from Eqs. (\ref{k1})-(\ref{gap}) turns out to be similar to that of TDGL, except that the non-equilibrium integral term in Eq. (\ref{gap}) accelerates relaxation of $\psi(t)$ at $q_h\simeq 1$. In both cases  superconductivity is destroyed at $q_h=1$. 

\begin{figure}
\includegraphics[width=\columnwidth]{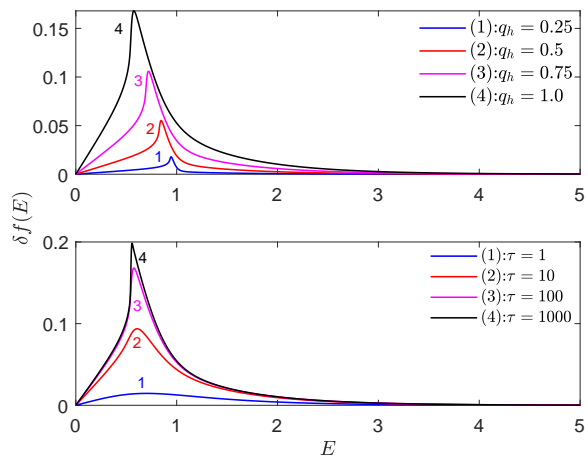}
\caption{The nonequilibrium correction $\delta f(E)$  at the times when the magnitude $\delta f(E,t)$ reaches maximum after the stepwise increase of $q(t)$. Taking $T=0.9T_c$, here the top panel shows $\delta f(E,t)$ calculated for different $q_h$ at $\tau=100$ and the bottom panel shows $\delta f(E,t)$ calculated for different values of $\tau$ at $q_h=1$.}
\label{fig10}
\end{figure}

Shown in Fig. \ref{fig10} are snapshots of a nonequilibrium part of the distribution function $\delta f(E,t)$ induced by the stepwise $q(t)$. Here the magnitude of $\delta f(E,t)$ calculated at $\tau=100$ increases as $q_h$ increases but remains relatively small up to $q_h=1$.  As the quasiparticle relaxation time $\tau$ increases, the magnitude of $\delta f(E,t)$ also increases. The peak in $\delta f(E,t)$ shifts to lower energies as $q_h$ increases, consistent with the decrease of the quasiparticle gap due to the dc current pairbreaking. 

\subsection{Fixed $Q(t)$.}
 
\begin{figure}
\includegraphics[width=\columnwidth ]{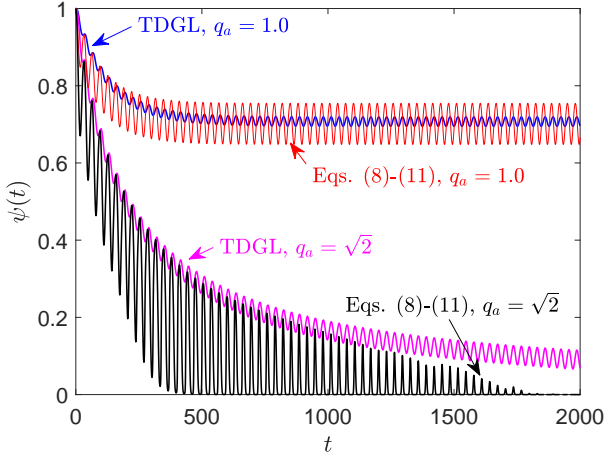}
\caption{Comparison of $\psi(t)$ calculated from the TDGL equations and Eqs. (\ref{k1})-(\ref{gap}) for $q(t)=q_a\sin\omega t$, $\tau=100$, $\omega=0.1$, and $T=0.9T_c$.}
\label{fig11}
\end{figure}

\begin{figure}
\includegraphics[width=\columnwidth ]{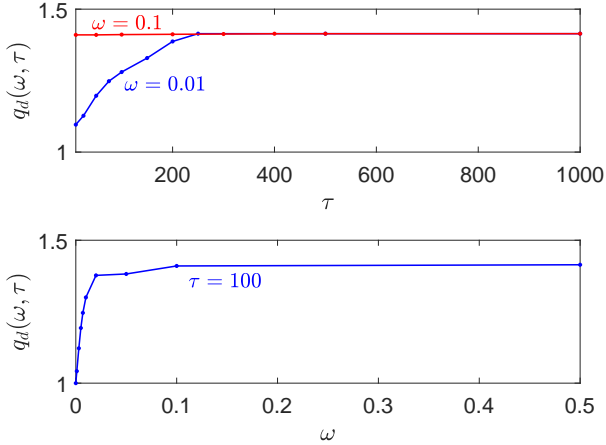}
\caption{Dynamic $q_d(\omega,\tau)$ as functions of $\tau$ (top) and $\omega$ (bottom) calculated from Eqs. (\ref{k1})-(\ref{gap}) at $T=0.9T_c$. }
\label{fig12}
\end{figure}

Solutions of Eqs. (\ref{k1})-(\ref{gap}) with $q(t)=q_a\sin\omega t$ are shown in Fig. \ref{fig11} along with the TDGL results obtained for the same input parameters. At $q_a=1$  the order parameters $\psi(t)$  oscillate around nearly the same mean values $\langle \psi\rangle$ but the amplitude of oscillations $\delta \psi(t)$ calculated from Eqs. (\ref{k1})-(\ref{gap}) is noticeably larger than the TDGL $\delta\psi(t)$. Relaxation of $\psi(t)$ from the initial value $\psi(0)=1$ to the steady-state oscillations described by Eqs. (\ref{k1})-(\ref{gap}) is also faster than the TDGL transient time, consistent with the above results for $q(t)=q_h\tanh t$ shown in Fig. \ref{fig9}.  These features become more pronounced at the dynamic critical momentum $q_d\simeq \sqrt{2}$ at $\omega\tau\gg 1$, where the amplitudes of oscillations $\delta\psi(t)$ grow significantly larger so that $\psi(t)$ touches zero but then recovers. Yet, despite a rather different dynamics of $\psi(t)$ described by Eqs. (\ref{k1})-(\ref{gap}) and the TDGL equations, superconductivity gets destroyed at the same critical value $q_d\to\sqrt{2}$ at $\tau=100$ and $\omega=0.1$ in both cases. The calculated dependencies of $q_d$ on $\tau$ and $\omega$ shown in  Fig. \ref{fig12} appear similar to the TDGL results shown by Fig. \ref{fig3}.

\begin{figure}
\includegraphics[width=\columnwidth ]{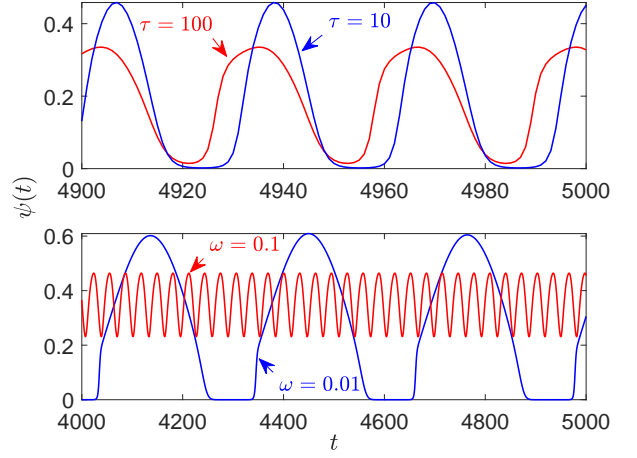}
\caption{Steady state oscillations of $\psi(t)$  calculated from Eqs. (\ref{k1})-(\ref{gap}) at $T=0.9T_c$ with $q=q_a\sin\omega t$ for: different $\tau$ at $\omega=0.1$ and $q_a=1.35$ (top) and different $\omega$ at $\tau=100$ and $q_a=1.30$ (bottom). }
\label{fig13}
\end{figure}

Our solutions of Eqs. (\ref{k1})-(\ref{gap}) have revealed a dynamic state in which $\psi(t)$ periodically vanishes but then recovers to $\psi(t)\sim 1$.  This state appears as the frequency decreases, as shown in Fig. \ref{fig13}. For instance, in the case of $\omega=0.1$ and $\tau=10$ shown in the top panel Fig. \ref{fig13}, $\psi(t)$ drops down to $\sim 2\times 10^{-3}$ at the minimum but remains finite. As $\psi(t)$ goes through the minimum the amplitude of $\delta f(E,t)$ decreases and changes sign. However,  at $\omega=0.01$ in the bottom panel, $\psi(t)$ at the minimum drops below the numerical tolerance level of $\sim 10^{-7}$ during a significant portion of the ac period. This case corresponds to a true transition to the normal state with $\psi=0$ in which all terms in Eq. (\ref{gap}) vanish and Eq. (\ref{k1}) describes an exponential relaxation of $\delta f(E,t)\propto \exp(-ts/2\tau)$ until the superconductivity recovers as $q(t)$ decreases. This behavior is physically transparent: at very low frequencies the quasi-static $\psi(t)$ is determined by the instantaneous $q(t)=q_a\sin\omega t$, resulting in periodic transitions to the normal state and the subsequent recovery of superconductivity once $|q(t)|$ exceeds $1$. At higher frequencies $\omega\gtrsim 0.1$, the superconducting state does not have enough time to disappear during the parts of the ac period in which $|q(t)|>1$, so that $\psi(t)$ at the minimum remains finite all the way to $q\to q_d$.    

\begin{figure}
\includegraphics[width=\columnwidth ]{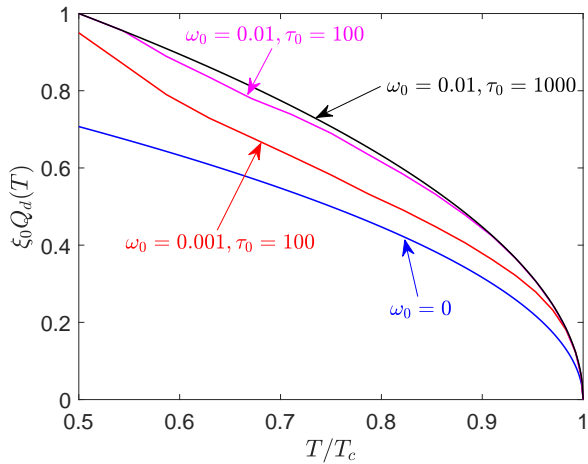}
\caption{$Q_d(T)$ calculated from Eqs. (\ref{k1})-(\ref{gap}) for different values of $\omega_0=\pi\hbar\Omega/8k_BT_c$ and $\tau_0=\tau_E(T_c)\Delta_0(0)$, where $\Delta_0^2(0)=8\pi^2T_c^2/7\zeta(3)$. The dynamic $Q_d=\sqrt{2(1-T/T_c)}/\xi_0$ at $\Omega\gg \Omega_c(T)$ has the same temperature dependence as the static $Q_c=\sqrt{1-T/T_c}/\xi_0$. If $\Omega\sim\Omega_c(T)$ the behavior of $Q_d(T)$ is affected by the temperature dependence of $\tau_E(T)$, as shown for the case of $\omega_0=0.001$ and $\tau_0=100$. }
\label{fig14}
\end{figure}

The calculated $Q_d(T)$ curves shown in   Fig. \ref{fig14} are similar to the TDGL results but generally fall below them: $Q_d(\Omega,T)\to \sqrt{2}Q_c=\sqrt{2(1-T/T_c)}$ at $\Omega\gtrsim\Omega_c(T)$ but $Q_d(\Omega,T)\to Q_c(T)$ at $\Omega\ll\Omega_c(T/2)$. The temperature dependence of $\tau(T)\propto T^{-3}$ results in a crossover of $Q_d(T,\Omega)$ from $Q_c(T)$ to $\sqrt{2}Q_c(T)$ as $T$ decreases. 

\subsection{Fixed $J(t)$}

\begin{figure}
\includegraphics[width=\columnwidth ]{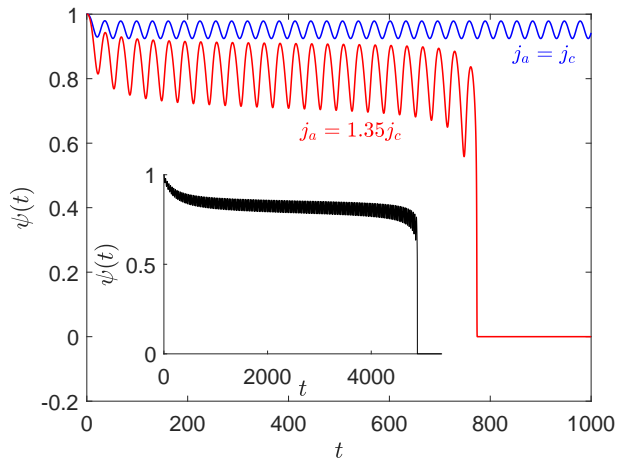}
\caption{Dynamics of $\psi(t)$ calculated at $j=j_a\sin \omega t$, $\omega=0.1$, $\tau=10$, $j_a=j_c$ and the critical current $j_a=1.35j_c$ at which $\psi(t)$ vanishes abruptly. The inset shows $\psi(t)$ calculated at $\tau=100$, $\omega=0.1$ and $j_a=\sqrt{2}j_c$. All calculations were performed at $T=0.9T_c$. }
\label{fig15}
\end{figure}

\begin{figure}
\includegraphics[width=\columnwidth ]{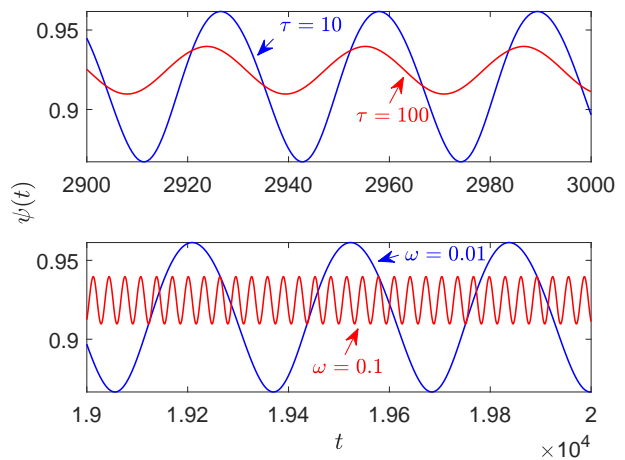}
\caption{ Steady state oscillations of $\psi(t)$ calculated at $T=0.9T_c$, $j=j_a\sin\omega t$,  $j_a=1.20j_c$ and: different $\tau$ at $\omega=0.1$ (top) and different $\omega$ at $\tau=100$ (bottom). }
\label{fig16}
\end{figure}

\begin{figure}
\includegraphics[width=\columnwidth ]{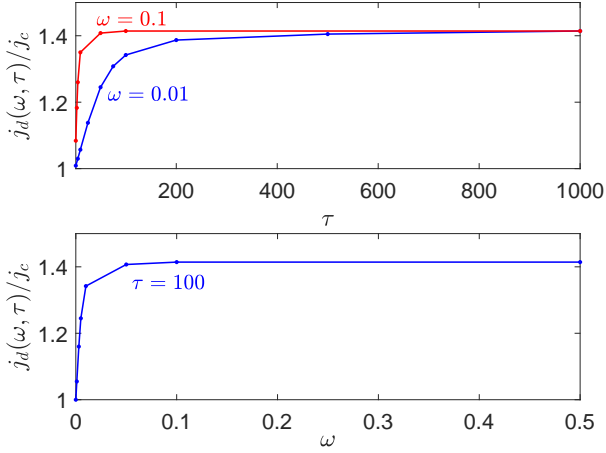}
\caption{Calculated dependencies of $j_d(\omega,\tau)$ on $\tau$ (top) and $\omega$ (bottom) at $T=0.9T_c$. Here $j_d$ levels off at $\sqrt{2}j_c$ at $\omega\tau\gtrsim 1$. }
\label{fig17}
\end{figure}

\begin{figure}
\includegraphics[width=\columnwidth ]{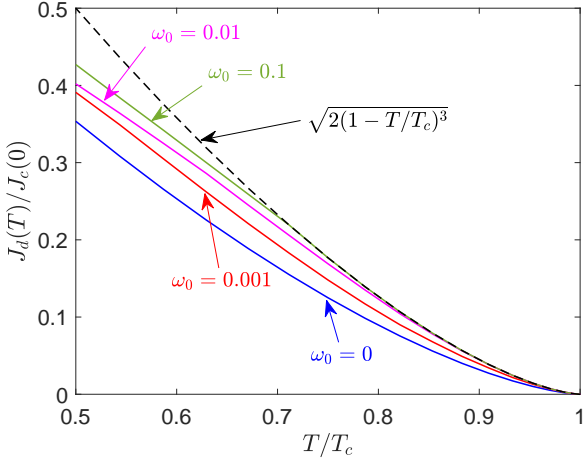}
\caption{$J_d(T)$ calculated from Eqs. (\ref{k1})-(\ref{usad})  for $\tau_0=100$ at different $\omega_0=\pi\hbar\Omega/8k_BT_c$ where $\tau_0=\tau_E(T_c)\Delta_0(0)$, and $\Delta_0^2(0)=8\pi^2T_c^2/7\zeta(3)$.  As $\Omega\gg \Omega_c(T)$, we have $J_d(T)=J_c(0)\sqrt{2}(1-T/T_c)^{3/2}$ at $T\to T_c$, however as $T$ decreases a crossover to $J_c(T)$ occurs even at $\Omega\geq\Omega_c(T)$.  }
\label{fig18}
\end{figure}

Solutions of Eqs. (\ref{k1})-(\ref{jeq}) for $j=j_a\sin\omega t$, $\omega=0.1$ at $\tau=10$ and $\tau=100$ shown in Fig. \ref{fig15} are qualitatively similar to that of $\psi(t)$ for a fixed $q(t)$. Here $\psi(t)$ vanishes abruptly at $j_a=j_d(\omega,T)$, the amplitude of oscillations of $\psi(t)$ essentially depends on $\omega$ and $\tau$, as shown in Fig. \ref{fig16}. The calculated $j_d=1.35j_c$ at $\tau(T)=10$ turned out to be slightly smaller than the TDGL value, but at $\tau(T)=100$  both TDGL theory and Eqs. (\ref{k1})-(\ref{jeq}) give the same $j_d=\sqrt{2}j_c$. The dependencies of $j_d(\omega,\tau)$ on $\tau$ and $\omega$ shown in Fig. \ref{fig17} appear similar to those for $q_d(\omega,\tau)$ in Fig. \ref{fig12} and clearly demonstrate that $j_d\to\sqrt{2}j_c$ at $\omega\tau\gg 1$. 
The temperature dependence of $J_d(\Omega,T)$ shown in Fig. \ref{fig18} is  similar to the TDGL results only at $T\to T_c$: $J_d(\Omega,T)\to\sqrt{2}J_c(0)(1-T/T_c)^{3/2}$ at $\Omega\gtrsim\Omega_c(T)$ and $J_d(\Omega,T)\to J_c(T)$ at $\Omega<\Omega_c(T)$. As $T$ decreases, the $J_d(\Omega,T)$ curves tend toward $J_c(T)$  even at $\Omega>\Omega_c(T)$.

\section{Nonlinear electromagnetic response}
\label{sec:man}

In this section we address an electromagnetic response of a nonequilibrium superconductor. For a nearly uniform current considered here, the linear response is quantified by a frequency-dependent complex conductivity,
\begin{equation}
\mathbf{J}=(\sigma_1-i\sigma_2)\mathbf{E},
\label{ohm}
\end{equation}
where $\sigma_1(\Omega)$ describes a dissipative quasiparticle response, $\sigma_2(\Omega)=1/\mu_0\Omega\lambda_L^2$ accounts for the Meissner effect, and $\lambda_L$ is the London penetration depth.  Here $\sigma_2$ also determines the kinetic inductance $\mathcal{L}_k=(d\Omega\sigma_2)^{-1}=\mu_0\lambda_L^2/d$ per unit length of a film of thickness $d$ ~\cite{kind1,kind2,kind3,kind4,kind5}. 
Using $\lambda_L^2(T)=2\hbar k_BT_c/\pi\mu_0\sigma_0\Delta_0^2$ near $T_c$~\cite{kopnin} yields:
\begin{equation}
\mathcal{L}_k=\frac{2\hbar k_BT_c}{\pi\sigma_0 d\Delta^2}.
\label{kinind}
\end{equation}

\begin{figure}
\includegraphics[width=\columnwidth ]{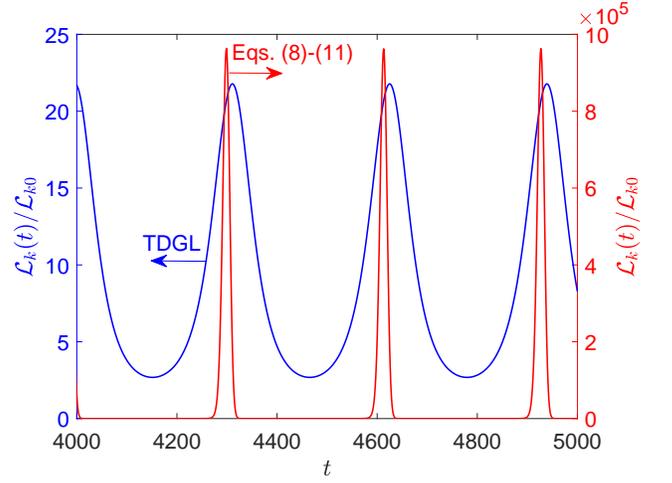}
\caption{Dynamics of $\mathcal{L}_k(t)$ in units of $\mathcal{L}_{k0}=\pi\sigma_0 d\Delta_0^2/2\hbar k_B T_c$ calculated from: (a) Eq. (\ref{glq}) and (b) Eqs. (\ref{k1})-(\ref{usad}) at $T=0.9T_c$ and $q(t)=q_a\sin\omega t$ with $q_a=0.9\sqrt{2}$, $\omega=0.01$, and $\tau=100$. Notice large-amplitude oscillations of $\mathcal{L}_k(t)$ at small $\omega\tau$ and large $q_a$, the peaks in $\mathcal{L}_k(t)$ calculated from Eqs. (\ref{k1})-(\ref{usad}) can be orders of magnitude larger than those obtained from Eq.(\ref{glq}).}
\label{fig19}
\end{figure}

At high current densities the conductivity $\sigma=\sigma_1-i\sigma_2$ depends on $Q(t)$, causing the nonlinear Meissner effect, intermodulation and generation of higher order harmonics of the electric field $E(t)$ in response to the ac current $J(t)=J_a\sin\Omega t$, ~\cite{Yip,Dahm,Anlage,Hirsch,Oates,Groll}.  
Defining the kinetic inductance by Eq. (\ref{kinind}), where $\Delta(t)$ is given by the solutions of Eqs. (\ref{glq}) or Eqs. (\ref{k1})-(\ref{gap}), we can expect strong oscillations of $\mathcal{L}_k(t)$ at large $J_a$ due to the nonequilibrium current pairbreaking. Shown in Fig. \ref{fig19} is the dynamics of $\mathcal{L}_k(t)$ calculated at a fixed $q(t)=q_a\sin\omega t$ with $q_a=0.9\sqrt{2}$, $\omega=0.01$ and $\tau=100$.  Here the amplitudes of $\mathcal{L}_k(t)$ increase with $q_a$ and diverge at $q_a\to q_d$, the peaks in $\mathcal{L}_k(t)$ getting higher as $\omega\tau$ decreases.  Figure \ref{fig19} also shows that the amplitudes of $\mathcal{L}_k(t)$ calculated from the full Eqs. (\ref{k1})-(\ref{usad}) can be orders of magnitude higher as compared to the TDGL results. This reflects larger amplitudes of oscillations of $\psi(t)$ calculated from Eqs. (\ref{k1})-(\ref{usad}) and discussed above (see Fig. \ref{fig11}). 

\begin{figure}
\includegraphics[width=\columnwidth ]{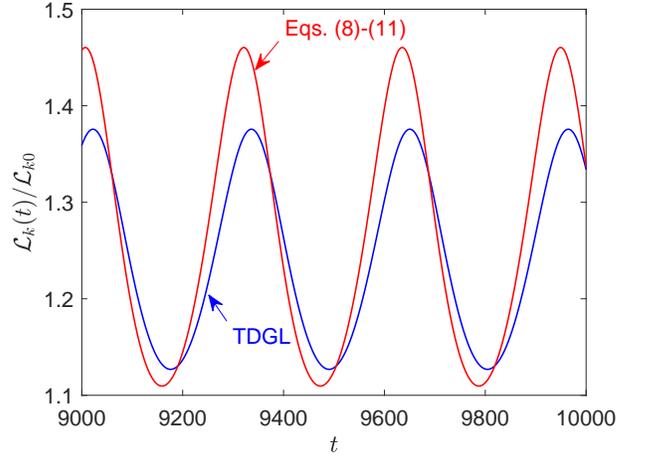}
\caption{Dynamics of $\mathcal{L}_k(t)$ calculated for a fixed current $j(t)=j_a\sin\omega t$ with $j_a=0.9\sqrt{2}j_c$, $\omega=0.01$ and $\tau=100$  using: (a) Eqs. (\ref{glq})-(\ref{glj}) and (b) Eqs. (\ref{k1})-(\ref{usad}) at $T=0.9T_c$. }
\label{fig20}
\end{figure}

Shown in Fig. \ref{fig20} is $\mathcal{L}_k(t)$ calculated from Eqs. (\ref{glq})- (\ref{glj}) and Eqs. (\ref{k1})-(\ref{usad}) at a fixed ac current $j=j_a\sin \omega t$ and $\tau=100$. Here $\mathcal{L}_k(t)$ can exhibit large-amplitude oscillations at small $\omega\tau$.  The amplitudes of $\mathcal{L}_k(t)$ calculated from  Eqs. (\ref{k1})-(\ref{usad}) are larger than the TDGL results, although not by orders of magnitude.

The above calculations of $\mathcal{L}_k(t)$ pertain to low frequencies $\omega\tau\ll 1$ at which $\mathcal{L}_k(t)$ follows instantaneously to the 
time-varying order parameter $\Psi(t)$. Generally, the nonlinear electromagnetic response at a fixed $q(t)=q_a\sin\omega t$ causes generation of multiple current harmonics:     
\begin{equation}
j(t)=\sum_{n}[j_{1n}\sin\omega_nt+j_{2n}\cos\omega_n t].
\label{jq}
\end{equation}
Likewise, the ac current $j=j_a\sin\omega t$ produces multiple harmonics of the electric field $\varepsilon=\dot{q}$:
\begin{equation}
\varepsilon(t)=\sum_{n}[\varepsilon_{1n}\sin\omega_nt+\varepsilon_{2n}\cos\omega_nt].
\label{qj}
\end{equation}
Here the frequencies $\omega_n$ and the Fourier amplitudes $j_{1n}(q_a)$, $j_{2n}(q_a)$, $\varepsilon_{1n}(j_a)$ and $\varepsilon_{2n}(j_a)$ are to be calculated self-consistently from Eqs. (\ref{k1})-(\ref{jeq}), as shown below.

\subsection{Fixed $q(t)$.}

Shown in Fig. \ref{fig21} are the current Fourier spectra calculated at different $\tau$ at $q_a=0.95\sqrt{2}$ and $\omega=0.1$. Here the multimode spectrum of $j(\omega)$ consisting of equidistant peaks at $\omega_n=n\omega$,  $n=1,3,5,...$ changes markedly as $\tau$ increases and the amplitudes of high-frequency harmonics diminish. The latter is consistent with the results of the previous sections which showed that at $\omega\tau\gg 1$ the amplitude of oscillations of superfluid density responsible for the generation of higher harmonics diminishes and the fundamental harmonic in $j(t)$ dominates. Here the nonequilibrium effects described by Eqs. (\ref{k1})-(\ref{gap}) significantly increase the amplitudes of higher order harmonics as compared to the respective TDGL results.  

Of particular interest is the dependence of the in-phase and out-of-phase parts of the amplitude of the main harmonic $j_m(t)=j_1\sin\omega t+j_2\cos\omega t$ on $q_a$, where $j_2$ determines the mean dissipative power $p=\omega q_a j_2/2$.  Shown in Fig. \ref{fig22} are steady-state oscillations of $j(t)$ at $\tau=1$ and $\tau=100$. At $q_a=2^{-1/2}$ and $\tau=100$, the current response is nearly in-phase with $q(t)$ but at $\tau=1$ the current has dips when $q(t)$ is maximum. The latter comes from pairbreaking effects which mostly reduce the superfluid density and the supercurrent when $q(t)$ reaches maximum. This effect becomes more  pronounced for a larger amplitude $q_a=0.95\sqrt{2}$ represented in Fig. \ref{fig22}(b).  In this case $\psi(t)$ is much reduced during a considerable part of the ac period so $j_1\ll j_2$ and the current response becomes nearly ohmic. 

The dependencies of the in-phase $j_1(q_a)$ and out of phase $j_2(q_a)$ amplitudes of the current main  harmonic on $q_a$ are shown in Fig. \ref{fig23} at $\tau=1$ and $\tau=100$.  At $\tau=100$ the response current is mostly in-phase with $q(t)$ up to the critical $q_a\approx\sqrt{2}$, while at $\tau=1$, the out-of-phase part of $j_m(t)$ is essential and significantly increases with $q_a$ and the supercurrent decreases. 

\begin{figure}
\includegraphics[width=\columnwidth ]{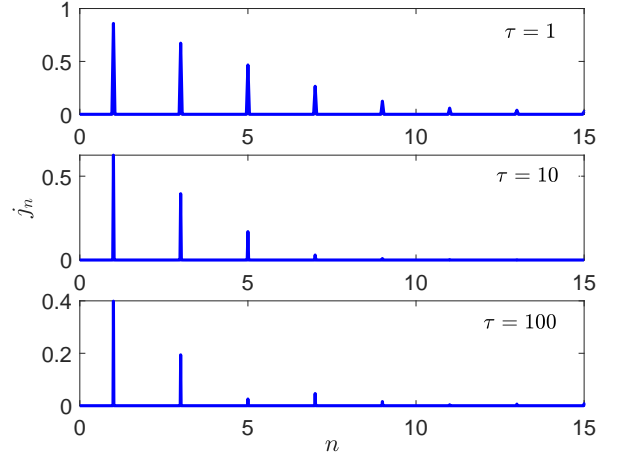}
\caption{ Fourier spectra of the current amplitudes $j_n=\sqrt{j_{1n}^2+j_{2n}^2}$ caused by $q(t)=q_a\sin\omega t$ calculated from Eqs. (\ref{k1})-(\ref{jeq}) for different $\tau$ at $T=0.9T_c$, $q_a=0.95\sqrt{2}$ and $\omega=0.1$. The Fourier amplitudes are peaked at $\omega_n=n\omega$ with $n = 1,3,5,..$.}
\label{fig21}
\end{figure}

\begin{figure}
\includegraphics[width=\columnwidth ]{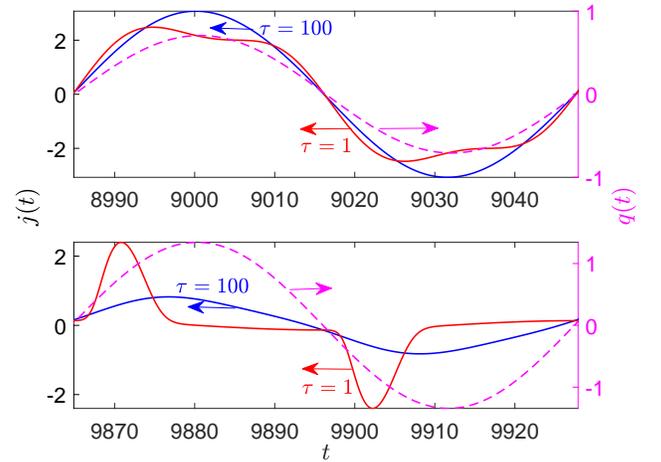}
\caption{ Nonlinear current response $j(t)$ calculated at $q_a=0.5\sqrt{2}$ and $q_a=0.95\sqrt{2}$ for two values of $\tau=1$ and $\tau=100$ at $T=0.9T_c$. At $\tau=100$ the current is nearly in phase with $q(t)$ at all $q_a$'s. At $\tau=1$ the current response at large $q_a$ becomes almost evenly divided into the in phase and out of phase parts.}
\label{fig22}
\end{figure}

\begin{figure}
\includegraphics[width=\columnwidth ]{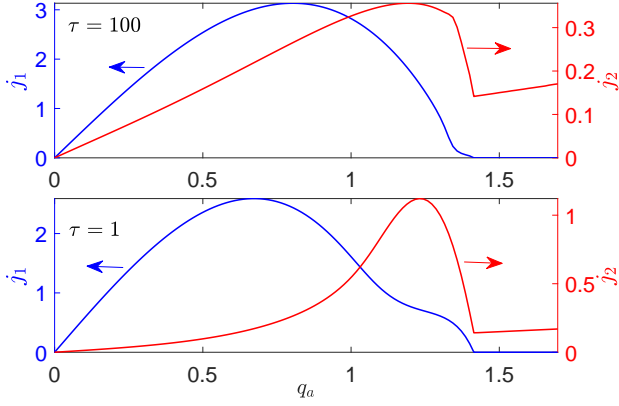}
\caption{ The amplitudes $j_1(q_a)$  and $j_2(q_a)$ of the main current harmonic as functions of $q_a$ calculated from Eqs. (\ref{k1})-(\ref{jeq}) at $T=0.9T_c$ with $q(t)=q_a\sin\omega t$ at $\omega=0.1$, $\tau=1$ and $\tau=100$. }
\label{fig23}
\end{figure}

\subsection{Fixed $j(t)$.}

\begin{figure}[htb]
\includegraphics[width=\columnwidth ]{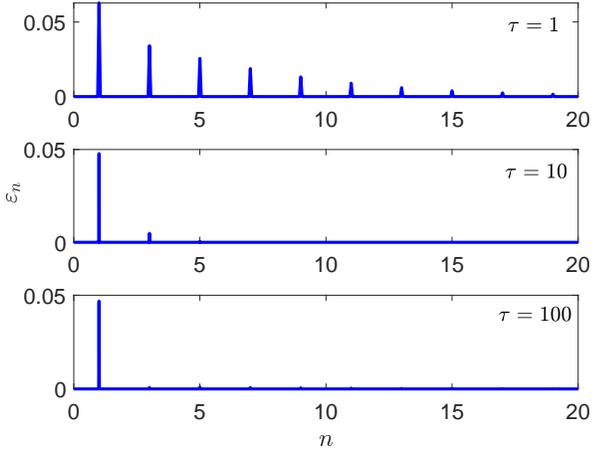}
\caption{ Fourier spectra of the electric field $\varepsilon_n=\sqrt{\varepsilon_{1n}^2+\varepsilon_{2n}^2}$ in response to the ac current $j=j_a\sin \omega t$ calculated from Eqs. (\ref{k1})-(\ref{jeq}) at $T=0.9T_c$, $j_a=0.77\sqrt{2}j_c$, $\omega=0.1$ and different $\tau$. The peaks in $\varepsilon_n$ occur at the odd multiples of $\omega$. }
\label{fig24}
\end{figure}

To calculate the Fourier harmonics of the dimensionless electric field $\varepsilon(t)=E(t)/E_0=\partial q/\partial t$ with $E_0=(2e\xi\tau_{GL})^{-1}$, we solved Eqs. (\ref{k1})-(\ref{jeq}) for $\psi(t)$ and $q(t)$ at a fixed ac current $j=j_a\sin\omega t$. Shown in Fig. \ref{fig24} are the Fourier spectra $\varepsilon(\omega)$ at $j_a=0.77\sqrt{2}j_c$, $\omega=0.1$ and different $\tau$. Like in the case of a fixed $q(t)$, the Fourier spectra of the electric field contain equidistant peaks at $\omega_n=n\omega$ with $n=1,3,5, ...$, the amplitudes of higher order harmonics decreasing as $\tau$ increases. 

\begin{figure}
\includegraphics[width=\columnwidth ]{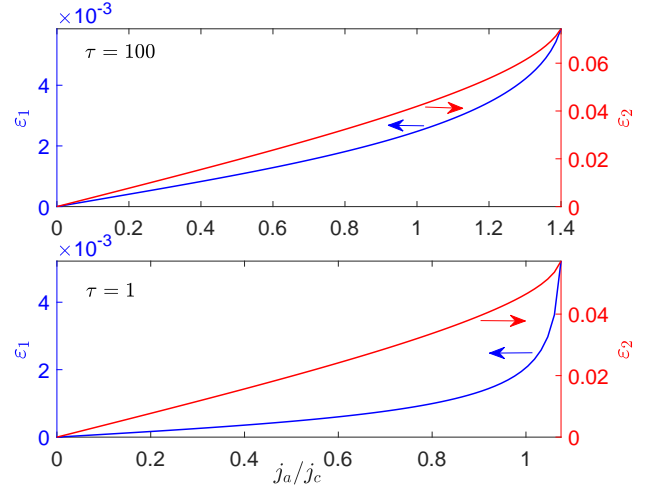}
\caption{ The amplitudes $\varepsilon_1(j_a)$ and $\varepsilon_2(j_a)$ of the main electric field harmonic as functions of $j_a$ calculated from Eqs. (\ref{k1})-(\ref{jeq}) at $T=0.9T_c$ with $j(t)=j_a\sin\omega t$, $\omega=0.1$ and $\tau=1$ and $\tau=100$. }
\label{fig25}
\end{figure}

\begin{figure}
\includegraphics[width=\columnwidth ]{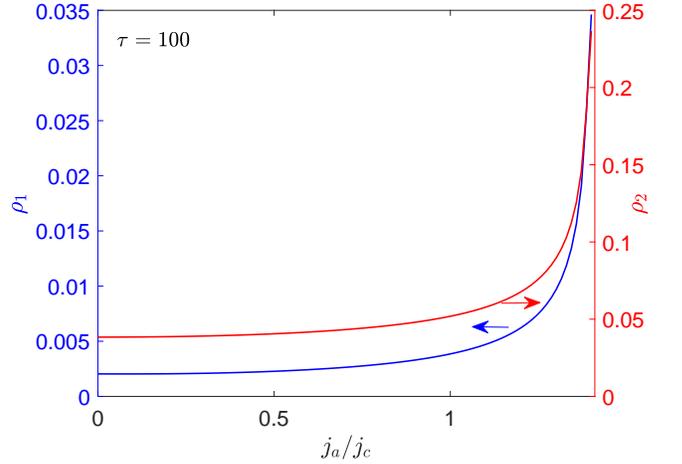}\\
\includegraphics[width=\columnwidth ]{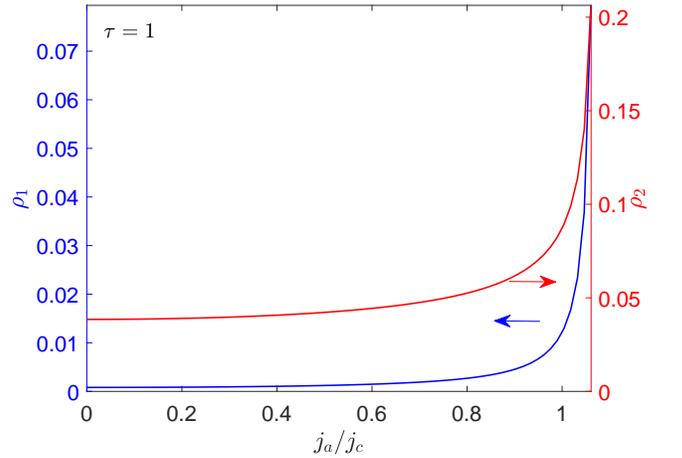}
\caption{ Differential resistivities $\rho_1$ and $\rho_2$ as functions of $j_a$ calculated from Eqs. (\ref{k1})-(\ref{jeq}) at $T=0.9T_c$ with $j(t)=j_a\sin\omega t$ at $\omega=0.1$, $\tau=1$ and $\tau=100$. }
\label{fig26}
\end{figure}

\begin{figure}
\includegraphics[width=\columnwidth ]{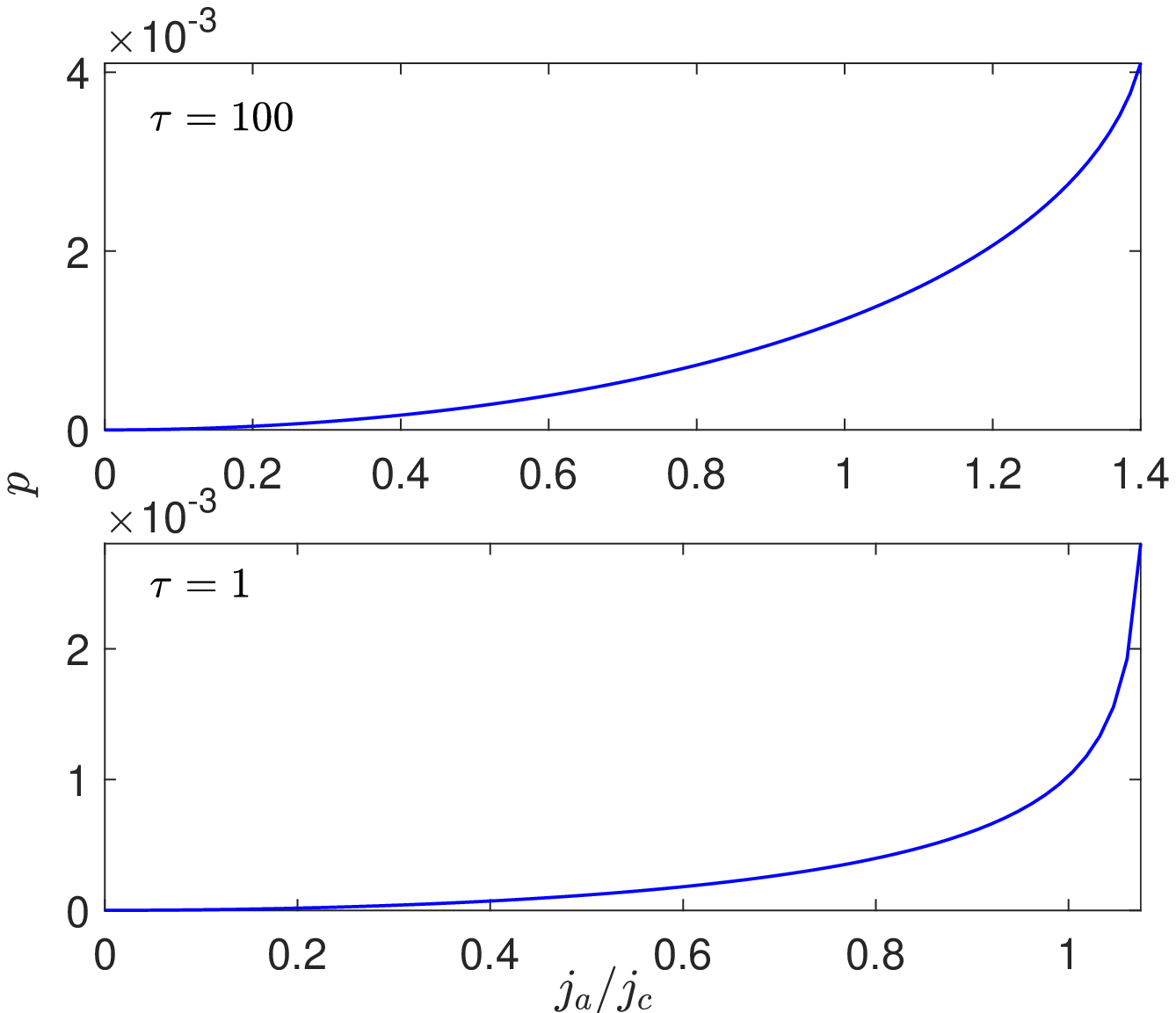}
\caption{ Ac power $p=\varepsilon_1 j_a/2$ as functions of $j_a$ calculated from Eqs. (\ref{k1})-(\ref{jeq}) at $T=0.9T_c$ with $j(t)=j_a\sin\omega t$ at $\omega=0.1$ for $\tau=1$ and $\tau=100$. }
\label{fig27}
\end{figure}

Figure \ref{fig25} shows the in-phase and out-of-phase amplitudes $\varepsilon_1$ and $\varepsilon_2$ of the main harmonic $\varepsilon_m(t)=\varepsilon_1\sin\omega t+\varepsilon_2\cos\omega t$ as functions of $j_a$ at $\omega=0.1$ and two values of $\tau=1$ and $\tau=100$. Here $\varepsilon_2(j_a)$ describing the superfluid response dominates at all $j_a$ and is nearly linear in $j_a$, indicating that the dynamic differential resistivity $\rho_2 =\partial \varepsilon_2/\partial j_a$  is weakly dependent on $j_a$ except for a sharp increase in a narrow region at $j_a\to j_d$ for both $\tau=1$ and $\tau=100$. By contrast, $\varepsilon_1(j_a)$ is linear in $j_a$ at $j_a\lesssim j_d/2$ but then  increases sharply as $j_a$ approaches $j_d$.  The differential resistivities $\rho_1(j_a)=\partial\varepsilon_1/\partial j_a$ and $\rho_2(j_a)=\partial\varepsilon_2/\partial j_a$ as well as the resulting dissipated power $p=P/P_0=\varepsilon_1 j_a/2$ as functions of $j_a$ where $P_0=E_0J_0$ are shown in Figs. \ref{fig26} and \ref{fig27}, respectively. At $J>J_d$ the supercurrent density vanishes jumpwise, resulting in the ohmic response $J=\sigma_0 E$ in the normal state. Notice that both $\rho_1$ and $\rho_2$ turned out to be much smaller than the normal state resistivity $\rho_0=1/\sigma_0$ in the whole region of $0<J_a<J_d$.

\section{Discussion}
\label{sec:disc}

In this work we address the breakdown of superconductivity by strong rf currents at $\hbar\Omega\ll \Delta_0\ll k_BT_c$. Here the deviation of the quasiparticle distribution function $f(E,t)$ from equilibrium is controlled by the amplitude of rf current and the inelastic electron-phonon scattering time $\tau_E$ which can be much larger than $\tau_{GL}$ and the rf period, $\Omega\tau_E\gg 1$. Because Eqs. (\ref{k1})-(\ref{jeq}) are applicable at $\hbar\Omega\ll k_BT_c$ ~\cite{LO,Kr1,Kr2}, they do not describe a microwave stimulation of superconductivity which occurs at $\hbar\Omega\gtrsim k_BT$ ~\cite{eliashberg}.  Yet the kinetic equations  (\ref{k1})-(\ref{jeq}) in which $\partial f/\partial E$ is replaced with its equilibrium value $\partial f_0/\partial E$ for a weak rf field ~\cite{Kr1,Kr2} can have spurious solutions corresponding to stimulated superconductivity. We did observe these solutions of the linearized Eqs. (\ref{k1})-(\ref{jeq}) but only at large rf amplitudes producing unphysical $\delta f(E,t)>1$. The results presented above are obtained using the Larkin-Ovchinnikov form of Eqs. (\ref{k1})-(\ref{jeq}) which include the exact $\partial f/\partial E$ ~\cite{LO}. In this case the nonequilibrium correction $\delta f(E,t)$ was always smaller than $1$ and no stimulated superconductivity was observed. 

The temperature and frequency dependencies of $Q_d$ and $J_d$ calculated from either the TDGL equations or Eqs. (\ref{k1})-(\ref{jeq}) turned out to be similar. Namely, both $Q_d$ and $J_d$ tend to their respective static GL values at $\Omega\tau_E\ll 1$ and gradually increase with frequency, approaching the universal values $Q_d\to \sqrt{2}Q_c$ and $J_d\to \sqrt{2}J_c$ at $\mbox{max}(\tau_{GL},\tau_E)\Omega\gg 1$. The physics of this effect is rather transparent: at $\Omega\tau_E\gg 1$, the pair potential $\psi(t)=\langle\psi\rangle+\delta\psi(t)$ undergoes small-amplitude rapid oscillations of $\delta\psi(t)$ around a mean value $\langle\psi\rangle$ which is determined by quasi-static equations with the time-averaged $\langle Q^2\rangle=Q_a^2/2$. Thus, the solutions for the mean order parameter $\langle\psi\rangle$ disappear above the same pairbreaking critical value of $\langle Q^2\rangle$ as for a dc current. This result can also be used to evaluate the dynamic superheating field $H_d$ at which the Meissner state in a large-$\kappa$ superconductor becomes absolutely unstable: 
\begin{gather}
H_d(T)\to H_s(T),\qquad \Omega\tau_E(T)\ll 1,
\label{hd1}\\
H_d(T)\to \sqrt{2}H_s(T),\qquad \Omega\tau_E(T)\gg 1,
\label{hd2} \\
H_s(T)=\left(\frac{\sqrt{5}}{3}+\frac{0.545}{\kappa} \right)H_c,\qquad \kappa\gg 1
\label{hc}
\end{gather}
where $H_s(T)$ is the dc superheating field at $T\approx T_c$  ~\cite{transtrum}.  At $\kappa\gg 1$ the screening current density varies slowly over $\xi$, so $Q(x,t)$ and $\Delta(x,t)$ are nearly independent  of the coordinate $x$ perpendicular to the surface.

The relation between the dynamic superheating field $H_d(T)$ and the dc superheating field $H_s(T)$ at low temperatures $T\ll T_c$ and frequencies $\hbar\Omega\ll k_BT_c$ has not yet been calculated from a microscopic theory. Yet based on the known dependence of the quasiparticle gap $\epsilon_g$ on the mean free path at $H=H_s$ ~\cite{lin}, we can make qualitative conclusions ~\cite{ags} regarding the essential effect of impurities on $H_d(T)$ at $T\ll T_c$. In the dirty limit $l\ll\xi_0$ at $T\ll T_c$, the quasiparticle gap $\epsilon_g(H)$ diminishes as the field increases but remains finite all the way to $H_s$ at which $\epsilon_g(H_s)\approx 0.38\Delta_0$ ~\cite{lin}, where $H_s=0.84H_c$~\cite{galaiko}.  In this case the density of thermally-activated quasiparticles remains exponentially small $n_{qp}(T)\lesssim n_0(\Delta_0/k_BT)^{1/2}\exp(-\epsilon_g/k_BT)$ in the entire field range of stability of the Meissner state, $0<H<H_s$. A low frequency field $\hbar\Omega\ll\Delta_0$  can produce nonequilibrium dquasiparticles which can affect dissipative kinetic coefficients and the surface resistance~\cite{ags}, but the effect of an exponentially small density of quasiparticles at $T\ll T_c$ on the dynamics of the superconducting condensate would be negligible, unlike the case of $T\approx T_c$ considered in this work. As a result, the condensate at $T\ll T_c$ reacts nearly instantaneously to the rf field with $\Omega\ll \Delta_0/\hbar$, despite slow kinetics of sparse quasiparticles, so the superconductivity would be destroyed under the same pairbreaking condition as in the absence of quasiparticles.  Thus, the dynamic superheating field $H_d$ of a dirty superconductor at $\hbar\Omega\ll\Delta_0$ and $T\ll T_c$ may be close to the static superheating field $H_s\approx 0.84 H_c$ 
even if $\Omega\tau_E\gg 1$.

For cleaner materials, the quasiparticle gap $\epsilon_g(H)$ vanishes before the dc depairing limit $H=H_s$ or $J=J_c$ is reached if $l\gtrsim 8.7\xi_0$~\cite{lin}. In this case the density of quasipartricles at $H=H_s$ is no longer negligible so their slow kinetics at $T\ll T_c$ may increase $H_d$ relative to $H_s$ even at $\hbar\Omega\ll \Delta_0$. A similar situation can also occur in superconductors with a nanostructured surface \cite{kg} or inhomogeneous density of impurities  \cite{sauls}, where the quasiparticle gap at the surface can be reduced by both the current pairbreaking and the proximity effect. Complex effects of impurities on the electron-phonon and electron-electron energy relaxation have been a subject of many experimental investigations  in recent years ~\cite{qu1,qu2,qu3,qu4}.   

Our calculations of a nonlinear electromagnetic response of a nonequilibrium superconducting state show that the amplitudes of higher order harmonics diminish as the quasiparticle energy relaxation time $\tau_E$ increases. Typically $\tau_E$ near $T_c$ is about 2 orders of magnitude higher than $\tau_{GL}$, except a narrow region of $T$ very close to $T_c$. Given that strong disorder can significantly reduce $\tau_E$ \cite{qu1,qu2,qu3,qu4}, one could expect that generation of higher order harmonics and intermodulation effects would be more pronounced in dirty superconductors. The moderate dependence of the dynamic differential resistivity $\rho_2(j_a)$ which defines a nonequilibrium kinetic inductance on $j_a$ shown in Fig. \ref{fig26} is qualitatively similar to that of ${ \cal L}_k(j_a)$ under the condition of the dc nonlinear Meissner effect \cite{Yip,Dahm,Hirsch,Groll}.  At the same time, the dissipative differential resistivity $\rho_1(j_a)$ shown in Fig. \ref{fig26} has a more pronounced dependence on $j_a$ than $\rho_2(j_a)$. Both $\rho_1(j_a)$ and $\rho_2(j_a)$ have strong peak as $j_a$ approaches the dynamic depairing current density but remain much smaller than the normal state resistivity at low frequencies $\hbar\Omega\ll \Delta$. The nonlinearity of $\varepsilon(j_a)$  in a nonequilibrium state manifests itself in a strong dependence of the rf dissipated power on the current amplitude, as shown in Fig. \ref{fig27}.  

\section*{Acknowledgments}

This work was supported by the US Department of Energy under Grant DE-SC0010081-020 and by the National Science Foundation under Grant PHY 1734075.

\appendix
%%%%%%%%%

\section{Nonequilibrium Equations} 
\label{Ap1}
The equations obtained in Refs. \onlinecite{ss,LO,Kr1,Kr2} for a nonequilibrium dirty s-wave superconductor at $T\approx T_c$ and $\Omega\ll \Delta_0$ include the quasi-stationary Usadel equation:
\begin{equation}
\!\!\frac{D}{2}\left[\alpha(\nabla-2ie\mathbf{A})^2\beta-\beta\nabla^2\alpha\right]=\left(\frac{1}{2\tau_E}-iE\right)\beta-\Psi\alpha,
\label{aus}
\end{equation}
where the normal and anomalous retarded Green's functions $\alpha(E)=N_1(E)+iR_1(E)$ and $\beta(E)=N_2+iR_2(E)$ satisfy $\alpha^2+\beta^2=1$.
Equation (\ref{aus}) is supplemented by the kinetic equations for the odd $f(E)$ and even $f_1(E)$ distribution functions of quasiparticles:
\begin{gather}
D\nabla\cdot\left[\left(N_{1}^{2}-R_{2}^{2}\right)\nabla\delta f\right]+2DN_{2}R_{2}\mathbf{Q}\cdot\left(\nabla f_{1}-e\frac{\partial f}{\partial E}\frac{\partial\mathbf{A}}{\partial t}\right)\nonumber \\-N_{1}\left(\frac{\partial}{\partial t}+\frac{1}{\tau_{E}}\right)\delta f=R_{2}\frac{\partial f}{\partial E}\frac{\partial|\Psi|}{\partial t},
\label{ak1}
\\
D\nabla\cdot\left[\left(N_{1}^{2}+N_{2}^{2}\right)\left(\nabla f_{1}-e\frac{\partial f}{\partial E}\frac{\partial\mathbf{A}}{\partial t}\right)\right]\nonumber \\+2DN_{2}R_{2}\mathbf{Q}\cdot\nabla\delta f-N_{1}\left(\frac{\partial}{\partial t}+\frac{1}{\tau_{E}}\right)\left(f_{1}+e\varphi\frac{\partial f}{\partial E}\right)\nonumber \\-N_{2}|\Psi|\left(2f_{1}+\frac{\partial f}{\partial E}\frac{\partial\theta}{\partial t}\right)=0,
\label{ak2}
\end{gather}
where $f=f_0+\delta f$ and $f_0=\tanh(E/2T)$.

The equations for $\Psi(\mathbf{r},t)=\Delta\exp(-i\theta)$ and ${\bf J}(\mathbf{r},t)$ are expressed in terms of $N_{1,2}$, $R_{1,2}$, $\delta f$ and $f_1$ as follows \cite{Kr1,Kr2}:
\begin{gather}
\left[\frac{\pi}{8T_c\epsilon}\frac{\partial}{\partial t}-\frac{1}{\Delta\epsilon}\int_0^{\infty}dE(R_2\delta f+iN_2 f_1)\right]\Psi= \nonumber \\
\xi^2(\nabla-2ie\mathbf{A})^2\Psi+\left(1-\frac{\Delta^2}{\Delta_0^2}\right)\Psi,
\label{aLO1} \\
\mathbf{J}=\frac{\pi\sigma_0}{4eT_c}\Delta^2\mathbf{Q}+\nonumber \\ \frac{\sigma_0}{e}\int_0^\infty dE\left[(N_1^2+N_2^2)\left(\nabla f_1-e\frac{\partial f}{\partial E}\frac{\partial\mathbf{A}}{\partial t}\right)+2N_2R_2\mathbf{Q}\delta f\right].
\label{aLOj}
\end{gather}
If $\delta f(E,r,t)$ and $\Psi(r,t)$ vary slowly over $\tau_E$, $\xi$ and $L_E=(D\tau_E)^{1/2}$, the derivatives in Eqs. (\ref{ak1})-(\ref{ak2}) can be neglected. In this  local equilibrium approximation  Eqs. (\ref{aus})-(\ref{aLOj}) reduce to Eqs. (\ref{gtdgl}) and (\ref{j}) ~\cite{Kr1,Kr2}.

If the spatial derivatives in Eqs. (\ref{aus})-(\ref{aLOj}) are negligible we readily obtain $f_1=-e\varphi\partial f/\partial E$ and $\Phi=-2e\varphi+\partial \theta/\partial t=0$ from Eq. (\ref{ak2}), giving $\nabla f_1-e(\partial f/\partial E)(\partial\mathbf{A}/\partial t)=1/2(\partial f/\partial E)(\partial \mathbf{Q}/\partial t)$. 
In turn, Eq. (\ref{aus}) reduces to the quartic equation:
\begin{gather}
\alpha^4-\mathcal{R}\alpha^3+\mathcal{S}\alpha^2+\mathcal{R}\alpha-\frac{\mathcal{R}^2}{4}=0,\nonumber \\
\mathcal{R} = \frac{2(u/\epsilon)^{1/2}(iE-1/2\tau)}{q^2}, \nonumber \\ 
\mathcal{S} = \frac{\mathcal{R}^2}{4}\left[\frac{\psi^2}{(iE-1/2\tau)^2}+1\right]-1,
\label{biquad}
\end{gather}
The relevant solution of Eq. (\ref{biquad}) is given by
\begin{equation}
 \alpha(E) = \frac{\mathcal{R}}{4}+\mathcal{E}+\frac{1}{2}\sqrt{-4\mathcal{E}^2-2\mathcal{A}-\frac{\mathcal{B}}{\mathcal{E}}},
 \label{solution}
\end{equation}
where
\begin{gather}
\mathcal{A} = \mathcal{S}-\frac{3\mathcal{R}^2}{8}, \qquad 
\mathcal{B} = 8\mathcal{R}+4\mathcal{R}\mathcal{S}-\frac{\mathcal{R}^3}{8}, \nonumber \\
\mathcal{C} = 2\mathcal{S}^3+27\mathcal{R}^2\mathcal{S}+27\mathcal{R}^2-\frac{27\mathcal{R}^4}{4}, \nonumber \\
\mathcal{D} = \left[\frac{1}{2}\left(\mathcal{C}+\sqrt{\mathcal{C}^2-4\mathcal{S}^6}\right)\right]^{1/3}, \nonumber \\ 
\mathcal{E} = \frac{1}{2}\sqrt{-\frac{2\mathcal{A}}{3}+\frac{1}{3}\left(\mathcal{D}+\frac{\mathcal{S}^2}{\mathcal{D}}\right)} \nonumber.
\end{gather}
%Equations (\ref{ak1}), (\ref{aLO1}) and (\ref{aLOj}) reduce to Eqs. (\ref{k1})-(\ref{jeq}).

\section{High-frequency limit, $\omega\tau\gg1$} 
\label{Ap2}
 
At high-frequencies  $\psi(t)=\psi+\delta\psi (t)$ has a small-amplitude oscillating component $\delta\psi(t)\ll \psi$ around 
a mean value $\psi$ so that $\langle\delta\psi\rangle=0$, where $\langle...\rangle$ denotes time averaging. In this case Eqs. (\ref{glq}) and (\ref{glj}) can be solved by the standard methods which have been developed for dynamic equations with rapidly oscillating parameters \cite{landau,bogoliubov}.  

\subsection{Fixed $Q(t)$.}

For a fixed $q(t)=q_a\sin\omega t$, we expand Eq. (\ref{glq}) up to quadratic terms in $\delta\psi$ and average over the rf period: 
\begin{gather}
r\dot{\psi}=\left(1-\langle q^2\rangle\right)\psi-\psi^{3}+\langle h\delta\psi\rangle-3\langle\delta\psi^{2}\rangle\psi,
\label{b2} \\
h(t)=\langle q^2\rangle-q^2(t)=\frac{q_a^2}{2}\cos2\omega t,
\label{b3}
\end{gather}
where $r=(1+4\tau^{2}\psi^{2})^{1/2}$, $\langle q^2\rangle=q_a^2/2$, and $\langle\delta\dot{\psi}\delta\psi\rangle=0$.

The dynamic equation for $\delta\psi(t)$ is obtained by expanding Eq. (\ref{glq}) up to linear terms in $\delta\psi$:
\begin{equation}
r\psi\dot{\psi}-g\delta\psi=h(t)\psi,\qquad g=1-q_a^{2}/2-3\psi^{2}
\label{b4}
\end{equation}
The solution of Eq. (\ref{b4}) is then:
\begin{gather}
\delta\psi(t)=A\cos2\omega t+B\sin2\omega t,
\label{b5} \\
A=-\frac{q_a^2g\psi}{2(4\omega^2r^2+g^2)},\qquad B=\frac{q_a^2\omega r\psi}{4\omega^2r^2+g^2}.
\label{b6}
\end{gather}
From Eqs. (\ref{b2}) and (\ref{b5}) we obtain the following self-consistency equation for $\psi(t)$:
\begin{equation}
r\dot{\psi}=\left(1-\frac{q_a^2}{2}\right)\psi-\psi^{3}+\frac{q_a^2A}{4}-\frac{3}{2}\psi (A^2+B^2).
\label{b7}
\end{equation}
At $4\omega^2r^2\gg g^2$, Eqs. (\ref{b6}) and (\ref{b7}) reduce to:
\begin{equation}
r\dot{\psi}=\left(1-\frac{q_a^{2}}{2}\right)\left(1-\frac{q_a^{4}}{32\omega^{2}r^2}\right)\psi-\psi^{3}.
\label{b8}
\end{equation}
Hence, the mean steady-state $\psi$ is given by:
\begin{equation}
\psi=\left(1-\frac{q_a^{2}}{2}\right)^{1/2}\!\!\left(1-\frac{q_a^{4}}{64\omega^{2}r^2}\right).
\label{b9}
\end{equation}
This state is stable with respect to small perturbations of $\psi(t)$ if $q_a<q_d=\sqrt{2}$. 
%As follows from Eq. (\ref{b9}),  
%there is no frequency correction to $q_d$ in leading order in $(4r\omega)^{-2}\ll 1$.

\subsection{Fixed $J(t)$.}

For a fixed $j(t)=j_a\sin\omega t$,
we linearize Eq. (\ref{glj}) with respect to an oscillating correction $\delta\psi(t)\ll 1$: 
\begin{equation}
j_a\sin\omega t=qu\psi^{2}+2u\psi q\delta\psi+\dot{q}.
\label{b11}
\end{equation}
Setting here $q(t)=q_{1}\sin\omega t+q_{2}\cos\omega t$ and $\delta\psi=A\cos2\omega t+B\sin2\omega t$, we obtain  
 $\langle q\delta \psi\rangle=0$, and $q(t)=-(j_a/u\psi^2)\sin\omega t$ in leading order in $\omega/u\ll1$ and $(\omega r)^{-2}\ll 1$. 
Substituting this $q(t)$ into Eq. (\ref{glq}) and averaging gives the equation for the mean $\psi(t)$: 
\begin{equation}
(1+4\tau^{2}\psi^{2})^{1/2}\dot{\psi}=\left(1-\frac{j_a^{2}}{2u^2\psi^{4}}\right)\psi-\psi^{3}.
\label{b12}
\end{equation}
The r.h.s. of Eq. (\ref{b12}) has the GL form for a fixed current except that the time averaging of $\langle q^2(t)\rangle=j_a^2/2u^2\psi^4$ reduces the current pairbreaking term in half as compared to the dc current.  As a result,
\begin{equation}
j_a^2=2u^2\psi^4(1-\psi^2),
\label{b13}
\end{equation}
Stability of the above steady state with respect to slow perturbations $\psi_{1}(t)$ can be addressed by setting $\psi(t)=\psi+\psi_1(t)$ and linearizing Eq. (\ref{b12})  with respect to $\psi_1$:
\begin{equation}
r\dot{\psi_{1}}=\left[1+\frac{3j_a^{2}}{2u^2\psi^{4}}-3\psi^{2}\right]\psi_1.
\label{b14}
\end{equation}
Hence, $\psi_1\propto \exp(\gamma t)$, where the decrement $\gamma$ is given by
\begin{equation}
\gamma=\frac{2}{r}\left(2-3\psi^2\right).
\label{b15}
\end{equation}
Here $j_a^2$ in Eq. (\ref{b14}) was expressed in terms of $\psi^2$ using Eq. (\ref{b13}). This state becomes unstable $(\gamma>0)$ at $j_{d}=\sqrt{2}j_c$ 
for which $j_a(\psi)$ reaches maximum at $\psi^2=2/3$.

\end{document}